\documentclass[12pt]{iopart}
\pdfoutput=1

\usepackage{amsfonts, amssymb, latexsym}
\usepackage{graphicx}
\usepackage{rotating}
\usepackage{color}
\usepackage[sort&compress,numbers]{natbib}
\usepackage{iopams}
\usepackage{xspace}
\usepackage[unicode,breaklinks]{hyperref}
\hypersetup{
    pdftitle={Dynamic Fisheye Grids for Binary Black Holes Simulations},
    colorlinks=false,            
    bookmarks=true,              
  }

\newcommand{\beq}[1]{\begin{equation} #1 \end{equation}}
\newcommand{\beqa}[1]{\begin{eqnarray} #1 \end{eqnarray}}

\newcommand{\pderiv}[2]{\frac{ \partial #1 }{ \partial #2 }}

\newcommand{\harm}{{\sc Harm3d}\xspace}   
\newcommand{\rin}{R_\mathrm{in}}
\newcommand{\rout}{R_\mathrm{out}}
\newcommand{\yin}{y_\mathrm{in}}
\newcommand{\yout}{y_\mathrm{out}}
\newcommand{\sech}{\mathrm{sech}}
\newcommand{\dF}{{^{^*}\!\!F}}

\newcommand{\bF}{{\bf F}}
\newcommand{\bU}{{\bf U}}

\newcommand{\sB}{\mathcal{B}}

\newcommand{\del}{{\partial}}
\newcommand{\bsq}{{{||b||^2}}}

\newcommand{\prim}{{{\mathbf{P}}}}

\newcommand{\dxp}[1]{{\Delta x^{\prime #1}}}
\newcommand{\xp}[1]{{x^{\prime #1}}}

\newcommand{\emf}[1]{{\mathcal{E}_{#1}}}
\newcommand{\dFn}{{^{^*}\!\!\hat{F}}}
\newcommand{\dFs}{{^{^*}\!\!\tilde{F}}}

\begin{document}

\title[]{Dynamic Fisheye Grids for Binary Black Holes Simulations}

\date{\today}

\author{Miguel Zilh\~ao, Scott C. Noble}

\address{Center for Computational Relativity and Gravitation and  School of Mathematical Sciences, Rochester Institute of Technology, Rochester, NY 14623, USA}

\eads{\mailto{mzilhao@astro.rit.edu}, \mailto{scn@astro.rit.edu}}

\begin{abstract}
We present a new warped gridding scheme adapted to simulating gas dynamics in binary black hole spacetimes. 
The grid concentrates grid points in the vicinity of each black hole to resolve the smaller scale structures there, and rarefies grid points
away from each black hole to keep the overall problem size at a practical level.  In this respect, our system can be thought of as a ``double'' version 
of the fisheye coordinate system, used before in numerical relativity codes for evolving binary black holes. 
The gridding scheme is constructed as a mapping between 
a uniform coordinate system---in which the equations of motion are solved---to the distorted system representing the spatial locations of our grid points. 
Since we are motivated to eventually use this system for circumbinary disk calculations, we demonstrate how 
the distorted system can be constructed to asymptote to the typical 
spherical polar coordinate system, amenable to efficiently simulating orbiting gas flows about central objects with little numerical diffusion.  
We discuss its implementation in the \harm code, tailored to evolve the magnetohydrodynamics (MHD) equations 
in curved spacetimes.   We evaluate the performance of the system's implementation in \harm with a series of tests, such as the advected magnetic field loop 
test, magnetized Bondi accretion, and evolutions of hydrodynamic disks about a single black hole and about a binary black hole. 
Like we have done with \harm, this gridding scheme can be implemented in other unigrid codes as a (possibly) simpler alternative to adaptive mesh refinement (AMR).
\end{abstract}




\section{Introduction}
\label{sec:intro}

When dealing with numerical simulations with a large range in length scales, it is often challenging to find the right balance between adequate resolution 
to accurately represent the relevant parts of the physical system and computational expense. 
Such a situation is typically encountered when evolving systems of compact objects (e.g. binaries involving neutron stars and black holes) in general relativity. 
In these cases, the stellar object's extent must be resolved by at most tenths of radii while the global spatial extent often must extend to $O(10^3)$ stellar 
radii or further in order evolve other important aspects of the problem, e.g., gravitational radiation, an extended circumbinary disk.  

The large range in spatial scales 
often leads those in the field of general relativistic MHD (GRMHD) to resort to AMR techniques, 
where the spacing between grid points changes 
depending on the desired accuracy of the solution
in a specific region.
For instance, block structured AMR~\cite{Berger84,Berger89} is used to resolve  compact objects  with a hierarchy of nested blocks of cells of ever decreasing size, 
where the grid cell size covering the stellar objects can be orders of magnitude smaller than the extent of the simulation domain~\cite{carpet_web,Moesta:2013dna,Anderson:2006ay,2012PhRvD..85l4010E}.  
Other codes refine locally as needed~\cite{2005ApJ...635..723A}, while others use multiple coordinate patches 
to cover a domain in a way beneficial for the computation~\cite{Calabrese2003:excision-and-summation-by-parts,Zink:2007xn,Duez:2008rb,2009ApJ...691..482F,2013PhRvD..87f4023R}.  
Though not adapted to general relativity (GR) yet, unstructured moving mesh codes 
show promise for automatic, local mesh refinement that achieves relatively low advection errors by moving cells at their local fluid element's velocity~\cite{2010MNRAS.401..791S,2011MNRAS.418.1392P,2011ApJS..197...15D}.   

Each of these methods has its advantages and drawbacks.  For instance, block structured AMR codes 
often employ Cartesian grids that lead to excessive loss of angular momentum and diffusion for nearly spherical flows.  They also do not scale very well due to challenges 
in balancing the computational load of managing the often large grid hierarchy data structure~\cite{Anderson:2006ay,Palenzuela2012}.  Multi-patch systems preserve angular momentum 
well~\cite{Zink:2007xn}, yet still require one to track the moving compact objects with a block mesh hierarchy, thereby sharing the scaling drawbacks of the block AMR codes. 
On the other hand, moving mesh codes seem to run at rates of several factors slower than those of their uniform static grid counterparts~\cite{2011MNRAS.418.1392P}.

The main motivation for this work comes from our ongoing efforts towards modeling accretion flows around 
binary black holes~\cite{Noble:2012xz}.  
In order to provide the scientific community with adequate predictions of electromagnetic signatures of supermassive 
binary black holes, accurate dynamic simulations of their gaseous environment must be made.   
Since the circumbinary 
gas is expected to be hot enough to be mostly ionized, GRMHD methods are essential for describing the role MHD physics plays in the accretion process.  
The magnetorotational instability~\cite{BH91,BH98} is expected to be a principal driver of angular momentum transport and it---in part---leads to the turbulence 
that ultimately dissipates the free energy of orbital motion into radiation~\cite{NK09,2013ApJ...769..156S}.
In the near field regime when the black holes are 
separated by less than $100 GM/c^2$ ($M$ is the total mass of the binary), several calculations have been made in the past few years 
that have begun to provide early descriptions of these events~\cite{2010ApJ...715.1117B,Farris11,Bode12,2012PhRvL.109v1102F,Giacomazzo12}.    One aspect that remains 
lacking in these calculations is a satisfactory description of the conditions of the binary's environment prior to merger.  Our research focuses on 
arriving at more realistic initial conditions by evolving from larger separations and for longer periods of time.  We must then employ a GRMHD code that can 
accurately describe the turbulent gas in the disk and the material that falls to the black holes.   

Unfortunately, all the codes used by other groups to explore
this problem have used the previously mentioned block-structured AMR techniques in Cartesian coordinates.  
These methods typically lead to poor conservation of fluid angular momentum and excessive
dissipation at refinement boundaries.  These two effects alter the disk's
angular momentum evolution and thermodynamics in nontrivial ways. 

We present here an alternative solution, which uses a predetermined coordinate system to concentrate more grid points in the spatial regions that require them with 
the GRMHD code called \harm~\cite{Noble09}.  The methodology we employ is similar to that employed in other GRMHD codes for fixed spacetimes (e.g.,~\cite{GMT03,dVH03,Noble09}), 
but our system can be time dependent 
and track moving features in the solution.  For the case of following two orbiting black holes, 
our system closely resembles a set of two ``fisheye'' transformations, one about each black hole.
Fisheye coordinates employ a transition function (e.g., a numerical approximation of the Heaviside function) to connect a region of high resolution
to a region of low resolution, so that computational effort is focused on the region of interest~\cite{Baker:2000zh,Baker:2001nu,Baker:2001sf}.  Single, static fisheye coordinates 
have been used extensively in numerical relativity\footnote{Fisheye coordinates are also used for visualizing streams of textual information 
\cite{2012arXiv1206.3980G,2013arXiv1305.5566Y}}.   For instance, they have been used for the vacuum two-body problem 
\cite{Zlochower:2005bj,Campanelli:2005dd,Campanelli:2006gf,Campanelli:2006uy,Campanelli:2006fg,Campanelli:2006fy}, and a series 
of nested fisheye coordinates have been used for GRMHD simulations of core collapse~\cite{PhysRevD.74.104026}.  Besides our system being dynamic, another key 
distinction is that we can focus points about each black hole rather than have one focal region encompassing both.  This ability increases our computational efficiency by
reducing the number of points in the region between the two black holes, a volume that does not require as much resolution for our problem.  Further, the system 
asymptotes to spherical coordinates away from the black holes, which is helpful for conforming more closely to the geometry of the circumbinary disk thereby 
providing better resolution of the disk's angular momentum and MHD turbulence~\cite{2013ApJ...772..102H}.  

Even though \harm lacks AMR (and, as mentioned above, AMR may not be best route), 
it has always supported arbitrary, time-dependent coordinate distortions. The coordinate transformation we  present here thus concentrates grid cells in the vicinity 
of the black holes and on the plane of the disk.   It also rarefies the number of cells as we move away from the binary so that the overall number of cells remains at 
a manageable level.   We again emphasize that, even though all our discussion will happen in the context of our implementation of this coordinate 
system in the \harm code, there is nothing in its construction that relies on \harm, or on MHD evolutions.  
The coordinate transformation we will present can in principle be successfully implemented in any (sufficiently modular) 
unigrid code and could be a simpler alternative to AMR.

The paper is presented in the following way. In Section~\ref{sec:theor-backgr} we cover some theoretical background, including some details about the MHD evolution procedure. 
In Section~\ref{sec:warp-cart-coord} we introduce our principal ideas by describing the construction of a distorted coordinate system adapted to 2-d periodic Cartesian grids;  we also 
describe how well this coordinate system performs with the advected magnetic field loop test. In Section~\ref{sec:warp-sph-coord} we state our warped spherical coordinate transformation, and 
show examples of possible grid configurations.  Results obtained with the ultimate spherical construction are illustrated in Section~\ref{sec:results}, where we demonstrate its
performance in evolving magnetized Bondi accretion, and hydrodynamic disks about single and binary black hole systems.  We end with some final remarks in Section~\ref{sec:final}.

\section{Theoretical Background}
\label{sec:theor-backgr}

Before we begin our description of the coordinate system itself, let us first discuss how coordinates are often used when solving equations of motion (EOM)
in the theory of GR, to provide a context.  Given two sets of coordinate systems, $x^\mu$ and $x^{\mu^\prime}$, that are related by a differential map, 
i.e.\  $x^\mu = x^{\mu}(x^{\mu^\prime})$ and $x^{\mu^\prime} = x^{\mu^\prime}(x^{\mu})$ are both continuous and differentiable, general covariance 
of GR implies that 
\beq{
\nabla_\mu T^{\mu \nu} \ = \ 0 \  = \ \nabla_{\mu^\prime} T^{\mu^\prime \nu^\prime} \quad , \label{covariance-of-conservation-of-stress-energy}
}
where $\nabla_{\mu}$ and $\nabla_{\mu^\prime}$ 
each represent the differential operator associated with the metric in their own system:  
\beq{
\nabla_{\mu} g_{\nu \lambda} \ = 0 \ = \nabla_{\mu^\prime} g_{\nu^\prime \lambda^\prime} \quad , \label{differential-operator}
}
and $\left(T_{\mu \nu}, g_{\mu \nu}\right)$ are related to $\left(T_{\mu^\prime \nu^\prime}, g_{\mu^\prime \nu^\prime}\right)$ via the well-known tensor transformation equation,
which we state here for completeness:
\beq{ 
{A_{\mu_1 \cdots \mu_{n1}}}^{\nu_1 \cdots \nu_{n2}} = {A_{\mu_1^\prime \cdots \mu_{n1}^\prime}}^{\nu_1^\prime \cdots \nu_{n2}^\prime} 
\prod_{i=1}^{n1}\prod_{j=1}^{n2} \pderiv{x^{\mu^\prime_i}}{x^{\mu_i}} \pderiv{x^{\nu_i}}{x^{\nu^\prime_i}} \,.
 \label{general-coordinate-transformation}
}
Greek indices represent spacetime coordinate indices.

The EOM are often solved numerically by discretizing the continuum domain into points and solving the equations for the fields at these points.  
One therefore must have a map from the computer memory's space to coordinate space of one's problem.   Even for the simple example of uniform Cartesian coordinates, a diffeomorphism is used 
to connect the memory's index space, $x^{\mu^\prime} = i^{\mu^\prime} \in \mathbb{Z}$, to $x^\mu$:  $x^\mu = x^\mu_0 + x^{\mu^\prime} dx^\mu$
with the grid spacing being $\Delta x^\mu = \pderiv{x^\mu}{x^{\mu^\prime}} \Delta x^{\mu^\prime} = dx^\mu$ since  $x^{\mu^\prime} \in \mathbb{Z}$\footnote{In the code, 
we promote $\left\{x^{\mu^\prime}\right\}$ to double precision floating point numbers because we need to represent intermediate coordinates, e.g., $x^{\mu^\prime}+\Delta x^{\mu^\prime}/2$.}.  
Often this map is constructed in a nonlinear way so that more grid points cover areas of interest.  Sometimes, one chooses to discretize w.r.t.\ the nonuniformly spaced 
$x^{\mu}$ and solve for $\left(g_{\mu \nu}, T_{\mu \nu}\right)$ (e.g.,~\cite{dVH03}), or one chooses to discretize w.r.t.\ the uniform $x^{\mu^\prime}$ system and solve for
the transformed quantities of interest $\left(g_{\mu^\prime \nu^\prime}, T_{\mu^\prime \nu^\prime}\right)$~\cite{Baker:2001sf,Noble09}.   The complication introduced by the former method is that finite differences
must be performed in a way valid for nonuniform grids, while the latter scheme often requires one to either transform the Christoffel symbols for the EOM's source terms
(or extrinsic curvature in the case of Einstein's equations), 
or finite difference the transformed metric $g_{\mu^\prime \nu^\prime}$ for them.  In passing, we note that the latter method is used in \harm and we choose to finite difference the metric
for the Christoffel symbols, primarily because deriving closed form expressions for the Christoffel symbols (transformed or otherwise) from our approximate binary black hole metric is 
impractical~\cite{Noble:2012xz,Mundimetal}; this step, therefore, costs no additional overhead for our scheme. 

\subsection{MHD Evolution}
\label{sec:mhd-evolution}

All our tests will involve the evolution of magnetized or unmagnetized matter, so we must introduce the EOM used by 
\harm.  The equations will also illustrate how the coordinates are relevant to the EOM we will be using. 

We assume that the gas does not self-gravitate and alter the spacetime
dynamics, so we need only solve the GRMHD equations on a specified background
spacetime, $g_{\mu \nu}(x^\lambda)$.  As mentioned earlier, \harm solves the EOM in uniform $x^{\lambda^\prime}$ coordinates
so please interpret the following tensor indices to be ``primed.''  

The EOM originate from the local
conservation of baryon number density, the local conservation of stress-energy, and the
induction equations from Maxwell's equations (please see~\cite{Noble09} for more
details).  They take the form of a set of conservation laws:
\begin{equation}
\del_t \bU\left(\prim\right) = 
-\del_i \bF^i\left(\prim\right) + \mathbf{S}\left(\prim\right) \, 
\label{conservative-eq}
\end{equation}
where $\bU$ is a vector of ``conserved'' variables, $\bF^i$ are the fluxes, 
and $\mathbf{S}$ is a vector of source terms.  Explicitly, these 
are 
\begin{eqnarray}
\bU\left(\prim\right) & = \sqrt{-g} \left[ \rho u^t ,\, {T^t}_t 
+ \rho u^t ,\, {T^t}_j ,\, B^k  \right]^T \label{cons-U} \\
\bF^i\left(\prim\right) & = \sqrt{-g} \left[ \rho u^i ,\, {T^i}_t + \rho u^i ,\, {T^i}_j ,\, 
\left(b^i u^k - b^k u^i\right) \right]^T \label{cons-flux} \\
\mathbf{S}\left(\prim\right) & = \sqrt{-g} 
\left[ 0 ,\, 
{T^\kappa}_\lambda {\Gamma^\lambda}_{t \kappa} ,\, 
{T^\kappa}_\lambda {\Gamma^\lambda}_{j \kappa}  ,\, 
0 \right]^T \, \label{cons-source}
\end{eqnarray}
where $g$ is the determinant of the metric, ${\Gamma^\lambda}_{\mu \kappa}$ are
the Christoffel symbols, $B^\mu = \dF^{\mu t}/\sqrt{4\pi}$ is our magnetic
field (proportional to the field measured by observers traveling orthogonal to the
spacelike hypersurface), $\dF^{\mu \nu}$ is the Maxwell tensor, $u^\mu$ is the
fluid's $4$-velocity, $b^\mu = \frac{1}{u^t} \left({\delta^\mu}_{\nu} + u^\mu
  u_\nu\right) B^\nu$ is the magnetic $4$-vector or the magnetic field projected
into the fluid's co-moving frame, and $W = u^t / \sqrt{-g^{tt}}$ is the fluid's
Lorentz function.  The MHD stress-energy tensor, $T_{\mu \nu}$, is defined as
\begin{equation}
T_{\mu \nu} = \left( \rho h + \bsq \right) u_\mu u_\nu   + \left( p + \bsq / 2\right) g_{\mu \nu} - b_\mu b_\nu \label{mhd-stress-tensor}
\end{equation}
where $\bsq \equiv b^\mu b_\mu$ is the magnetic energy density, $p$ is the gas pressure,
$\rho$ is the rest-mass density, $h = 1 + \epsilon + p/\rho$ is the specific
enthalpy, and $\epsilon$ is the specific internal energy. 

If not otherwise noted, we use the following numerical techniques. 
Piecewise parabolic reconstruction of the primitive variables is performed 
at each cell interface for calculating the local Lax-Friedrichs flux~\cite{GMT03}, and a 
3-d version of the FluxCT algorithm is used to impose the solenoidal constraint,
$\partial_i \sqrt{-g} B^i = 0$~\cite{2000JCoPh.161..605T}.  The EMFs
(electromotive forces) are calculated midway along each cell edge using
piecewise parabolic interpolation of the fluxes from the induction equation.  A
second-order accurate Runge-Kutta method is used to integrate the EOM using the
method of lines once the numerical fluxes are found.  The primitive variables
are found from the conserved variables using the ``2D'' scheme of~\cite{Noble06}.  
Please see~\cite{Noble09} and \ref{app:magn-fields-gener} for more details.


\section{Warped Cartesian Coordinates}
\label{sec:warp-cart-coord}

As mentioned in Section~\ref{sec:intro}, our goal is to construct a grid tailored to evolving accretion disks in the background of binary black hole spacetimes. For this purpose, in order to better resolve the accretion disk itself, a spherical coordinate system is desired.
As proof of principle and as a first test, however, we will start by presenting a 2-dimensional warped coordinate system adapted to a (2-d) periodic Cartesian grid between two coordinate systems: $\left\{T,X,Y\right\}$ and $\left\{t,x,y\right\}$.
This warped Cartesian coordinate system also serves a pedagogical purpose, since it is not as involved as the warped spherical construction that will follow.

\subsection{Implementation}
\label{sec:implementation}

We fix the following notation. The physical (spatial) coordinates are labeled $\left\{X,Y\right\}$, they are assumed to span the physical range $\left[X_{\rm min}, X_{\rm max}\right]$, $\left[Y_{\rm min}, Y_{\rm max}\right]$ and can have (in general) a non-uniform grid spacing.
The ``numerical'' coordinates (i.e., the coordinates actually used to evolve the equations in the numerical code) are labeled $\left\{x,y\right\}$, are uniformly discretized, and span $[0,1]$.

Our immediate goal is to develop a coordinate transformation between these two coordinate systems that can concentrate grid cells on certain (predetermined) physical regions. In general, we will be interested in having 
two ``focal points,'' where the density of cells is largest.

In the following, we summarize the desired properties of the coordinate transformation:
\begin{enumerate}
\item $T = t$ ; \label{prop:time-is-time}
\item $x , y  \in [0,1]$, $X  \in \left[X_{\rm min}, X_{\rm max}\right], Y \in \left[Y_{\rm min}, Y_{\rm max}\right]$;\label{prop:xy-XY-bounds}
\item $\Delta x, \Delta y = \mathrm{constant}$;\label{prop:const-dx}
\item $\left(X_1,Y_1\right)$ and $\left(X_2,Y_2\right)$  are the $\left(X,Y\right)$ coordinates of the two focal points, assumed to correspond to  $\left(x_1,y_1\right)$ and 
  $\left(x_2,y_2\right)$  in  $\left(x,y\right)$ coordinates;\label{prop:bh-pos}
\item $X_1, X_2, Y_1, Y_2, x_1, x_2, y_1, y_2$ are (generically) functions of time; \label{prop:time-funcs}
\item $\pderiv{X}{x}$ and $\pderiv{Y}{y}$  should have local minima along the lines
$\left\{\left( X_1 , Y \right), \left(X_2,Y\right)\right\}$ and $\left\{\left( X , Y_1 \right),\left(X,Y_2\right)\right\}$, respectively; \label{prop:dx-minima}
\item $\pderiv{X}{x}$ and $\pderiv{Y}{y}$ should be periodic in $x$ and $y$, respectively: $\pderiv{X}{x}(x) = \pderiv{X}{x}(x + 1)$, $\pderiv{Y}{y}(y) = \pderiv{Y}{y}(y + 1)$;\label{prop:dx-peridocity}
\item $\pderiv{X}{x} > 0 $, $\pderiv{Y}{y} > 0 $;\label{prop:dx-dy-positivity}
\end{enumerate}

Property~\ref{prop:dx-minima} guarantees that resolution is highest at the focal points. 
Property~\ref{prop:dx-dy-positivity} ensures there are no coordinate singularities.  The periodicity 
condition, \ref{prop:dx-peridocity}, makes our tests easier and will be used in the spherical case later on 
since the azimuthal coordinate has a similar symmetry.  

\begin{figure}[h]
  \centering
  \includegraphics[width=0.6\textwidth]{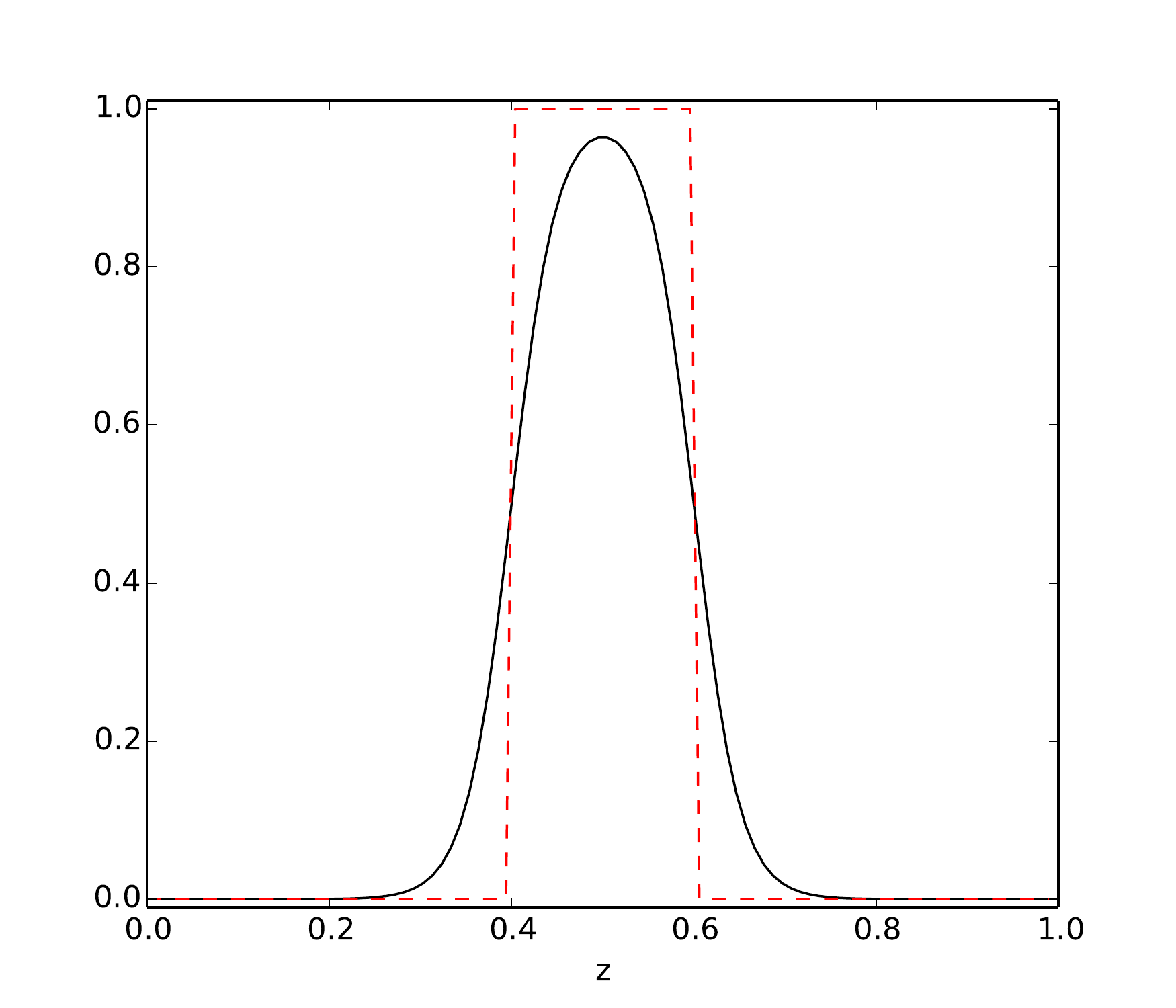}
  \caption{Ideal boxcar function [Eq.~(\ref{top-hat})] (dashed red line) and our approximate 
    version (solid black line), with $z_1 = 0.5$, $\delta=0.1$, $h=20$. \label{fig:top-hat} }
\end{figure}

The coordinate transformation satisfying these properties is most easily described 
using the following idealized functionals: 
\begin{eqnarray}
\sigma(z) & =  
\left\{ 
\begin{array}{ccc}
-1 &: &z < 0\\
1  &: &z \ge 0\\
\end{array}
\right.
\label{step-func} \\
\Sigma(z) & =  \int \sigma(z) dz = 
\left\{ 
\begin{array}{ccc}
-z  &: &z < 0\\
z   &: &z \ge 0\\
\end{array}
\right. 
\label{int-step-func} \\
\tau(z,z_1,\delta) & = \frac{1}{2}\left[ \sigma\left(z-z_1+\delta\right) - \sigma\left(z-z_1-\delta\right) \right]
\label{top-hat} \\
\mathcal{T}(z,z_1,\delta) & = \frac{1}{2}\left[ \Sigma\left(z-z_1+\delta\right) - \Sigma\left(z-z_1-\delta\right) \right] 
\label{int-top-hat} \\
\tilde{\tau}(z,z_1,\delta) & = \tau(z,\bar{z}_1,\delta) + \tau(z,\bar{z}_1-1,\delta) + \tau(z,\bar{z}_1+1,\delta)
\label{periodic-top-hat} \\
\tilde{\mathcal{T}}(z,z_1,\delta) & = \mathcal{T}(z,\bar{z}_1,\delta) + \mathcal{T}(z,\bar{z}_1-1,\delta) + \mathcal{T}(z,\bar{z}_1+1,\delta)
\label{periodic-int-top-hat}
\end{eqnarray}
where $\bar{z} = z - \mathrm{floor}(z)$. Expression~(\ref{top-hat}) (the ``boxcar'' function) will be used throughout, and can be seen schematically in Figure~\ref{fig:top-hat}. Tilded expressions are nothing but the periodic equivalents of their non-tilded counterparts.

We found that following system meets our requirements:
\begin{eqnarray}
 \frac{1}{X_{\rm max}-X_{\rm min}}\pderiv{X}{x} & = 1 
& - a_{x1} \ \tilde{\tau}\!\left(y,y_1,\delta_{y3}\right) \left[ \tilde{\tau}\left(x,x_1,\delta_{x1}\right) - 2 \delta_{x1} \right]  \nonumber\\
& {} & - a_{x2} \ \tilde{\tau}\!\left(y,y_2,\delta_{y4}\right) \left[ \tilde{\tau}\left(x,x_2,\delta_{x2}\right) - 2 \delta_{x2} \right]  \label{dX-dx-3} \\
 \frac{1}{Y_{\rm max}-Y_{\rm min}}\pderiv{Y}{y} & =  1 
& - a_{y1} \ \tilde{\tau}\!\left(x,x_1,\delta_{x3}\right) \left[ \tilde{\tau}\left(y,y_1,\delta_{y1}\right) - 2 \delta_{y1} \right]  \nonumber\\
& {} & - a_{y2} \ \tilde{\tau}\!\left(x,x_1,\delta_{x4}\right) \left[ \tilde{\tau}\left(y,y_2,\delta_{y2}\right) - 2 \delta_{y2} \right]  \label{dY-dy-3}
\end{eqnarray}
where the $a_{x,y}$ and $\delta_{x,y}$ are tunable parameters that approximately represent the 
relative amplitude and width of distortion, respectively. 

We here note, though, that in practice we will not be using the idealized expressions~(\ref{step-func},\ref{int-step-func}). Instead, we will use numerical approximations to the step function, its integral, 
and its derivative (respectively):
\begin{eqnarray}
\sigma(z,z_1,h) & = \tanh\left(h\left(z-z_1\right)\right) \\
\Sigma(z,z_1,h) & = \int \sigma(z,z_1,h) dz  = \frac{1}{h} \ln \cosh \left(h\left(z-z_1\right)\right) \\
\sigma^\prime(z,z_1,h) & = \partial_z \sigma(z,z_1,h) = h \, \sech^2\left(h\left(z-z_1\right)\right)
\end{eqnarray}
where $h$ is a steerable parameter controlling the ``steepness'' of the function's transition. The other functions are defined as before without any change.
To accommodate the fact that the approximate boxcar function does not have the same integral as the 
ideal boxcar function, we must normalize it differently in Eqs.~(\ref{dX-dx-3},\ref{dY-dy-3}) by making
the following change: $2 \delta_{x1} \rightarrow \tilde{\mathcal{T}}(1,x_1,\delta_{x1}) - \tilde{\mathcal{T}}(0,x_1,\delta_{x1})$, and analogously for $\delta_{x2}, \delta_{y1}, \delta_{y2}$.

This yields the final expressions for $X(x,y)$ and $Y(x,y)$:
\begin{eqnarray}
\fl \frac{X(x,y)-X_{\rm min}}{X_{\rm max}-X_{\rm min}} = x 
& - a_{x1} \ \tilde{\tau}\!\left(y,y_1,\delta_{y3}\right) \Big\{ & \, \tilde{\mathcal{T}}\!\left(x,x_1,\delta_{x1}\right) 
  -  \tilde{\mathcal{T}}\!\left(x_1,x_1,\delta_{x1}\right) \nonumber \\
&{}& - \left( x - x_1 \right) \left[ \tilde{\mathcal{T}}(1,x_1,\delta_{x1}) - \tilde{\mathcal{T}}(0,x_1,\delta_{x1}) \right] \, 
  \Big\} \nonumber \\
&{}- a_{x2} \ \tilde{\tau}\!\left(y,y_2,\delta_{y4}\right) \Big\{ & \, \tilde{\mathcal{T}}\!\left(x,x_2,\delta_{x2}\right) 
  -  \tilde{\mathcal{T}}\!\left(x_2,x_2,\delta_{x2}\right) \nonumber \\
&{}& - \left( x - x_2 \right) \left[ \tilde{\mathcal{T}}(1,x_2,\delta_{x2}) - \tilde{\mathcal{T}}(0,x_2,\delta_{x2}) \right] \, 
  \Big\} \label{X-eq-1} \\[0.5cm]
\fl \frac{Y(x,y)-Y_{\rm min}}{Y_{\rm max}-Y_{\rm min}} = y
& - a_{y1} \ \tilde{\tau}\!\left(x,x_1,\delta_{x3}\right) \Big\{ & \, \tilde{\mathcal{T}}\!\left(y,y_1,\delta_{y1}\right) 
  -  \tilde{\mathcal{T}}\!\left(y_1,y_1,\delta_{y1}\right) \nonumber \\
&{}& - \left( y - y_1 \right) \left[ \tilde{\mathcal{T}}(1,y_1,\delta_{y1}) - \tilde{\mathcal{T}}(0,y_1,\delta_{y1}) \right] \, 
  \Big\} \nonumber \\
&{} - a_{y2} \ \tilde{\tau}\!\left(x,x_2,\delta_{x4}\right) \Big\{ & \, \tilde{\mathcal{T}}\!\left(y,y_2,\delta_{y2}\right) 
  -  \tilde{\mathcal{T}}\!\left(y_2,y_2,\delta_{y2}\right) \nonumber \\
&{}& - \left( y - y_2 \right) \left[ \tilde{\mathcal{T}}(1,y_2,\delta_{y2}) - \tilde{\mathcal{T}}(0,y_2,\delta_{y2}) \right] \, 
  \Big\} \label{Y-eq-1}
\end{eqnarray}

\begin{figure}[htbp]
\centering
\includegraphics[width=0.7\textwidth]{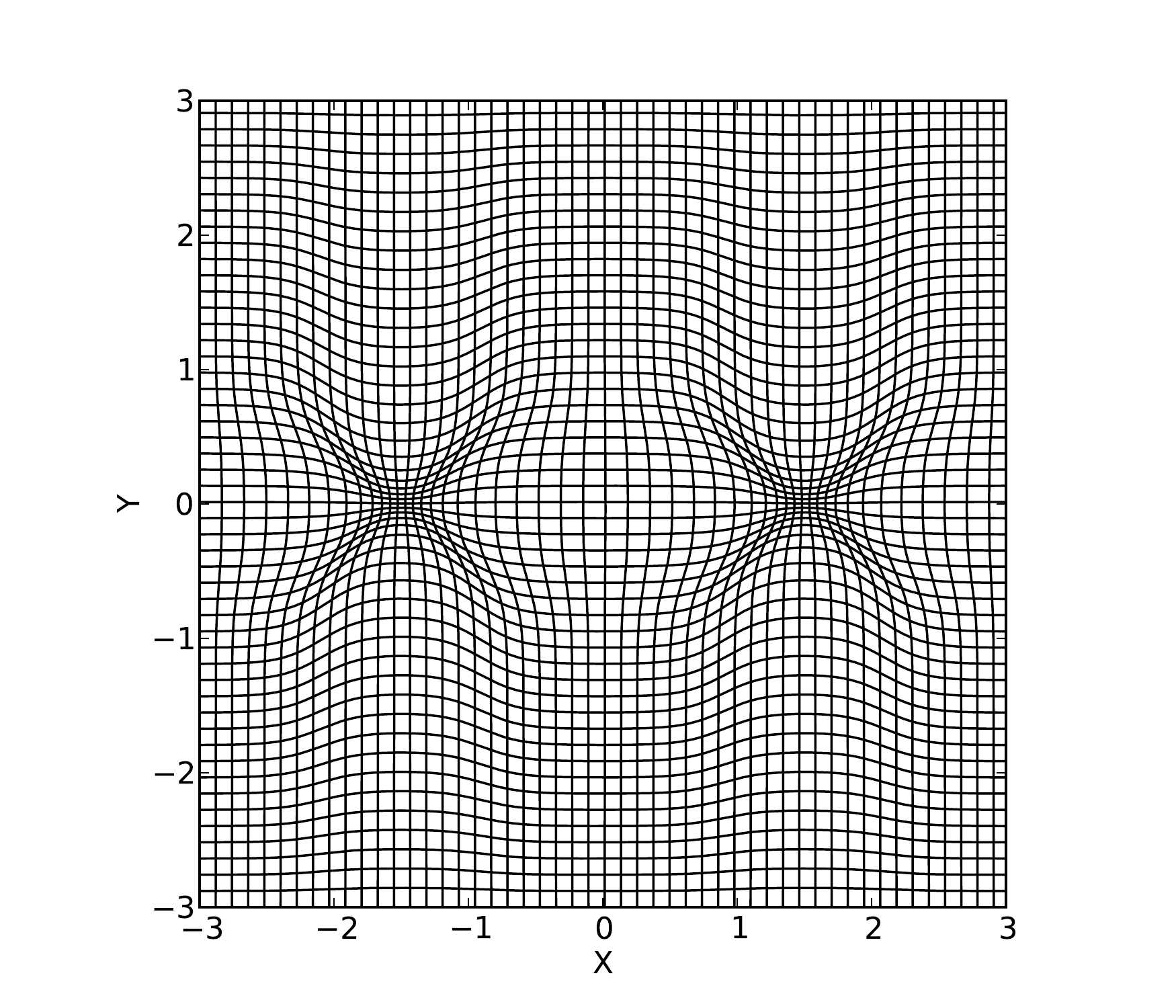}
\caption[]{Example grid with 
$\delta_{x1} = \delta_{x2} = \delta_{x3} = \delta_{x4} = \delta_{y1} = \delta_{y2} = \delta_{y3} = \delta_{y4} = 0.1$, 
$h_{x1} = h_{x2} = h_{y1} = h_{y2} = 20$, 
$a_{x1} = a_{x2} = a_{y1} = a_{y2} = 1$, 
$x_1 = 0.25$, $x_2=0.75$, $y_1 = y_2 = 0.5$, $X_{\rm max}=-X_{\rm min} = 3$, $Y_{\rm max}=-Y_{\rm min} = 3$. \label{fig:cart-grid} }
\end{figure}

With this coordinate system we have in general two focal points, $(x_1, y_1)$ and $(x_2, y_2)$, that have a larger density of grid cells in their vicinity. Parameters $a_{x,y}$ and $\delta_{x,y}$ control the transition to the outer region.
As stated before, the location of these focal points can vary in any prescribed continuous fashion, to better accommodate the dynamics of the physical system.  For the sake of clarity, we have dropped the inclusion of the $h$ 
parameters in the $\tau$ functions in the above expressions.  In general, there is an $h$ parameter for 
every $\delta$ parameter, i.e.\ we have  $h_{x1}$, $h_{x2}$, $h_{y1}$, and $h_{y2}$.   
The plethora of parameters illustrates the great control one has in fine-tuning a system of coordinates. 
We show in Figure~\ref{fig:cart-grid} an example with two focal points with symmetric distortions. 

\subsection{Advected Field Loop}
\label{sec:advected-field-loop}

As a first test of the performance of these coordinates---and validation of its implementation in \harm---we 
present results from the so-called ``advected field loop'' test, wherein one starts with a 
circular loop of magnetic field  at the center and constant gas density, pressure, and velocity throughout~\cite{DeVore:1991aa,Toth:1996aa,2005JCoPh.205..509G,2011ApJS..193....6B,Moesta:2013dna}.
Even though the exact solution is trivial (i.e.\ no evolution in the field except for the configuration's translation in space), 
the test is difficult for constrained transport (CT) schemes to evolve without the field significantly diffusing and deforming.  
Since the focal point moves through the domain, the coordinates change in time. 
One could imagine that the moving distortion could lead to additional diffusion or deformation of the field loop.   
Our aim here is to compare our results from using the distorted coordinates to what results from using uniform coordinates, all with \harm.  

We specify the initial conditions as follows. Working in the magnetic loop's rest frame, we specify the vector potential to be $A_{\mu} = \left(0, 0, 0, A_{z} \right)$, where
\begin{equation}
A_z=\left\{\begin{array}{rl}
A_{\rm loop} \left(R_{\rm loop}-r\right); & r \le R_{\rm loop}\,\,\\
0; & r > R_{\rm loop}\,\,
\end{array}\right. .
\end{equation}
We next perform a coordinate transformation on $A_{\mu}$ to the boosted frame with coordinate velocity $-u^i/u^t$ and then transform to the numerical coordinates 
$A_{\mu^\prime}$. We finally compute $B^{i^\prime}$ from $A_{\mu^\prime}$ 
through a finite difference procedure consistent with our CT scheme [Eq.~(\ref{divergence-less-stencil-condition})].  
We choose parameters $A_{\rm loop} = 10^{-3} , \rho = 1.0, P = 3.0$ for
the initial magnetic field and hydrodynamic configuration. 
The fluid is given uniform velocity throughout, $u^i/u^t = \left[1/12,0,0\right]$, and periodic boundary conditions are imposed at each edge.

\begin{figure}[tphb]
\centering
\includegraphics[width=0.7\textwidth]{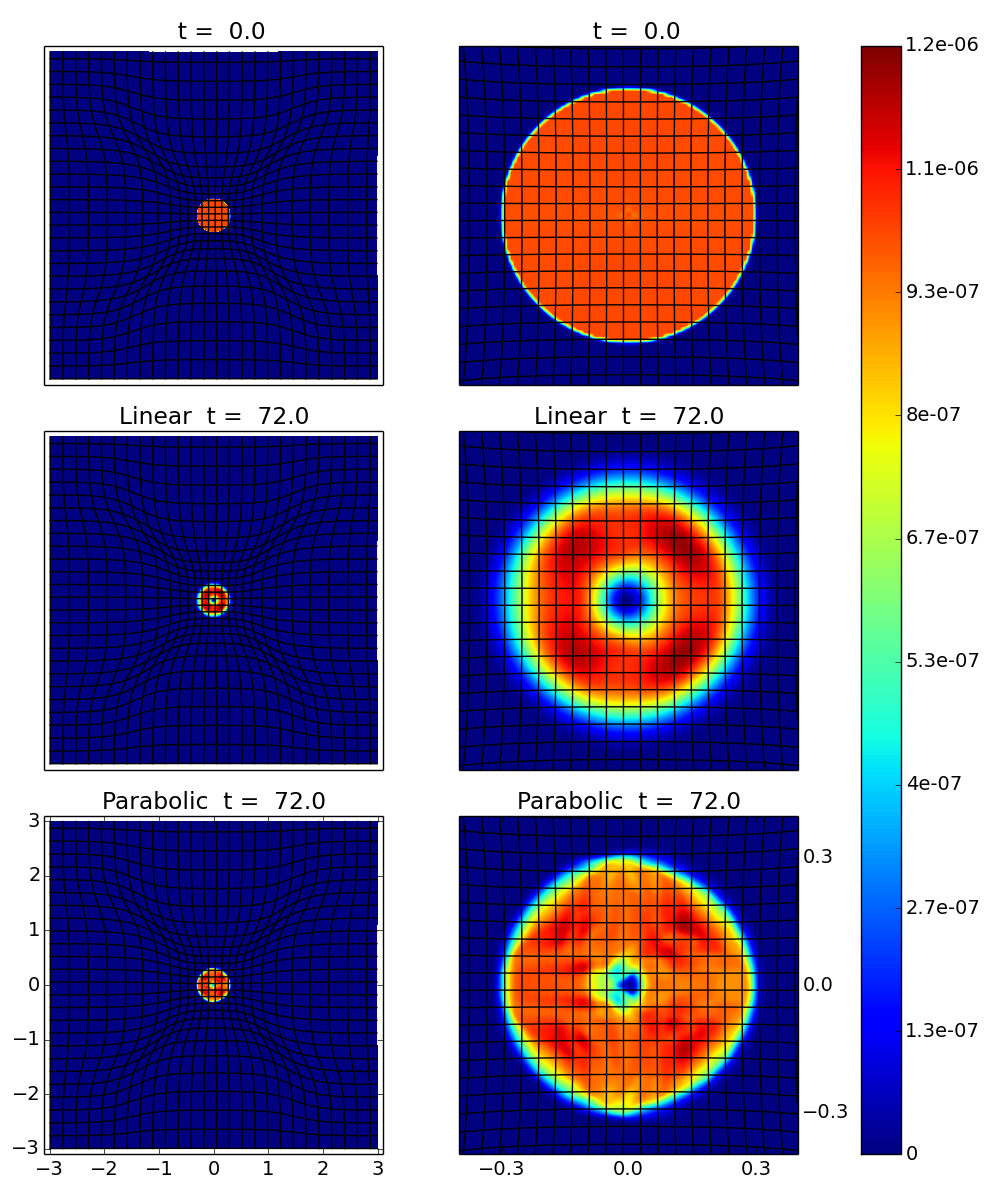}
\caption{Two-dimensional distributions of magnetic field energy density ($\bsq$) plotted at initial time (top row) and at the final time of the test (bottom two rows).  We show results from runs 
using the linear reconstruction method (middle row), and the parabolic reconstruction method (bottom row). 
In all runs, we used $384 \times 384$ cells.  The loop was advected  across the grid with a constant velocity along the $x$-axis, and the ``warp'' was transported at the same rate.  We show 
both the entire domain with 15 cells per grid line shown (left column), and a view focusing on the field loop with 5 cells per grid line shown (middle column).
Grid parameters used for this simulation were $\delta_{x1} = \delta_{x3} = \delta_{y1} = \delta_{y3} = 0.25$, $h_{x1} = h_{y1} = 12$, 
$a_{x1} = a_{y1} = 1$, $a_{x2} = a_{y2} = 0$, $X_{\rm max}=-X_{\rm min} = 3$, $Y_{\rm max}=-Y_{\rm min} = 3$.  The center of the warp, $\left(x_1,y_1\right)$, is set to coincide with the 
center of the loop as predicted by the advection equation.
\label{fig:loop-evol-horizontal}  }
\end{figure}

\begin{figure}[tphb]
\centering
\includegraphics[width=0.7\textwidth]{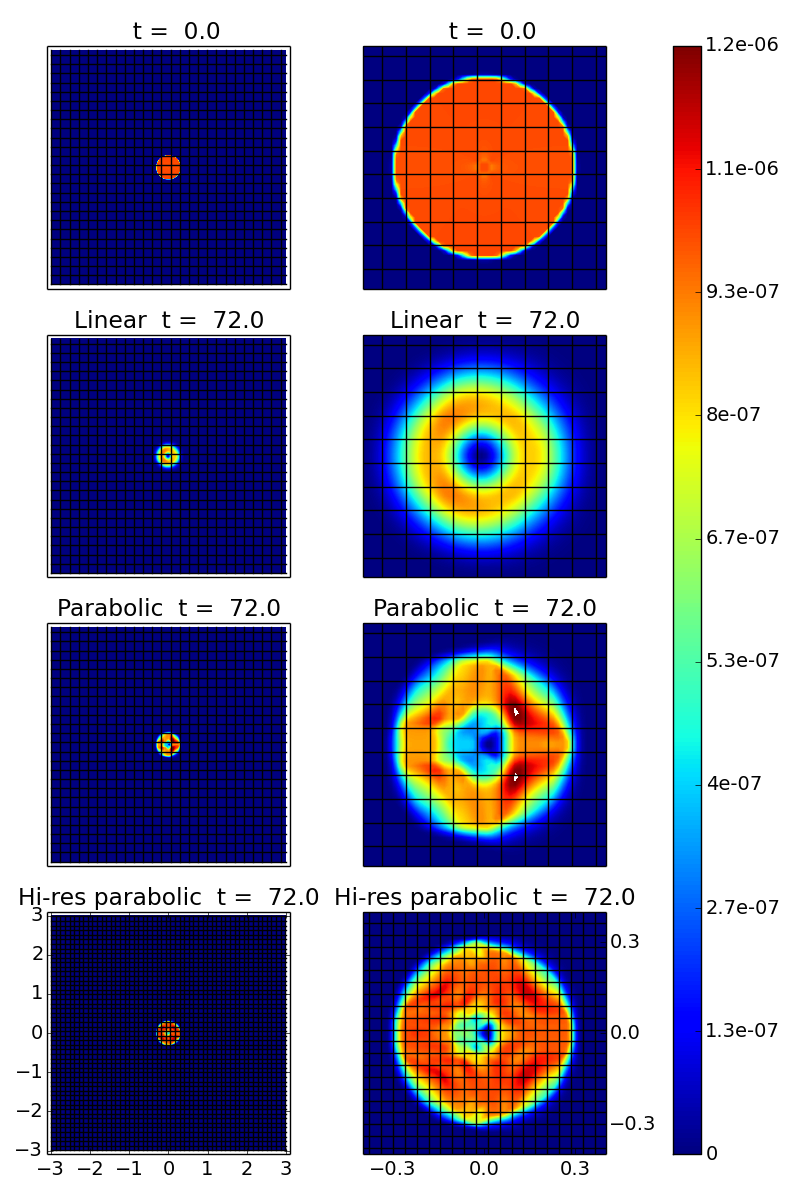}
\caption{Equivalent runs to those illustrated in 
Figure~\ref{fig:loop-evol-horizontal}, but using a static uniform grid with  $384 \times 384$  cells.  
We show both the entire domain with 15 cells per grid line shown (left column), and a view focusing on the field loop with 5 cells per grid line shown (right column). 
Bottom row shows a higher-resolution unigrid case ($768 \times 768$ cells). The grid spacing here was chosen to match the smallest
grid spacings found in the warped configurations.
\label{fig:loop-evol-cart-horizontal} }
\end{figure}

For this physical scenario, we set up the coordinate system with only one focal point, fix the parameters 
in a way such that the cells with smaller aspect ratio lie well outside of the loop, and make the local grid spacing
inside the loop resemble a uniform one, cf.\ Figure~\ref{fig:loop-evol-horizontal}.   
We evolve the system for one period, until $t=72$, and compare the results against the initial configuration 
at $t=0$. These results should be compared with those obtained with no warping, which we present in 
Figure~\ref{fig:loop-evol-cart-horizontal}.  We also explore the differences between two sets of numerical 
schemes used for reconstructing the EMFs at the cell edges for the CT scheme, and primitive variables at the cell interfaces.
One uses nearest neighbor (2-point) averaging for the EMFs and piecewise linear reconstruction (monotonized central limiter~\cite{1977JCoPh..23..276V})
for the primitive variables (the so-called ``linear'' method); the other uses a piecewise parabolic interpolation method~\cite{1984JCoPh..54..174C} for both the EMFs and primitive variables 
(the so-called ``parabolic'' method). Once reconstructed at the cell edges, the EMFs are used in the same 
way for the two FluxCT schemes, which makes both CT schemes second-order accurate even though the parabolic reconstruction often yields more accurate results; 
please see~\ref{app:magn-fields-gener} 
for more details regarding our FluxCT schemes.



\begin{figure}[htbp]
\centering
\includegraphics[width=0.6\textwidth]{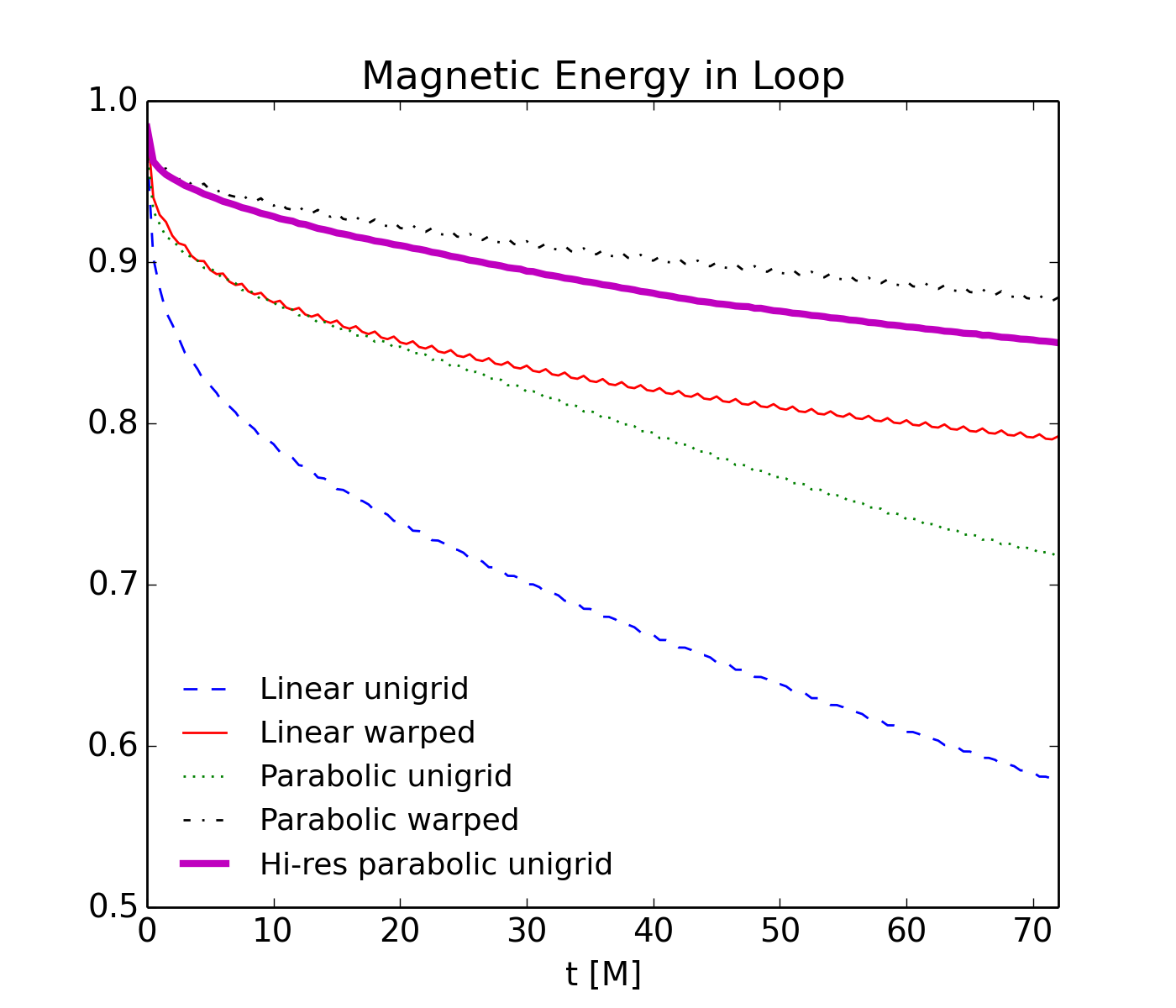}
\caption[]{Integral of $\bsq$ over the field loop as function of time [Eq.~(\ref{magnetic-energy-in-loop})] using a variety of run configurations. The 
standard linear method (solid and dashed) and the parabolic method  (dotted and dash-dotted) were used, both with a uniform Cartesian grid (dashed and dotted) and a 
warped mesh (solid and dash-dotted).  Another configuration included using the parabolic method with a higher-resolution uniform grid ($768 \times 768$ cells), whose grid spacing was chosen to match the smallest
grid spacings found in the warped configurations. \label{fig:bsq_diffusion} }
\end{figure}

As can be appreciated from these figures, some (inevitable) diffusion happens for all cases. The warped evolutions, 
however, while developing some asymmetry, seem to diffuse much less overall than their uniform counterparts.  We also find
that the evolutions using the parabolic method also diffuse less than those using the linear method.  The parabolic method 
seems to result in a higher effective resolution with the same number of cells.  On the other hand, the higher order 
interpolation stencils used by the parabolic method introduce short wavelength structure into the solution.  
The differences in rates of diffusion between the various methods can be quantitatively confirmed in 
Figure~\ref{fig:bsq_diffusion}, where we integrate the magnetic field's energy density over the initial 
field loop's area at each time step.  Specifically, we show 
\beq{
\frac{1}{A_{\rm loop}^2}\int_{r < R_\mathrm{loop}}   \bsq \sqrt{-g^\prime} \, dx^{\prime 1} \, dx^{\prime 2} \quad 
\label{magnetic-energy-in-loop}
}
as a function of time, where the radius $r$ used in the integral's limits is the 2-d radius from the loop's instantaneous center (as predicted by its advection velocity).
We note that the integral is performed w.r.t.\ numerical coordinates in both the uniform and warped cases. 
Diffusion of magnetic field leads to a loss of field within this area.  Similar to the effect of the warped coordinates, the parabolic method effectively increases 
the resolution and diminishes the rate of diffusion outside the loop.  The warped/linear combination, however, is even less diffusive than the uniform/parabolic combination.   Obviously, the warped/parabolic combination 
is the least diffusive of all the configurations, resulting in only a $10\%$ loss of magnetic field energy out of the loop after a period.   In order to measure the effective resolution of the warped system, we
performed a uniform Cartesian run with double the number of cells per dimension (i.e.\ $768 \times 768$ cells), as this is approximately the local resolution of the warped grid within the field loop.  We found that 
the warped system leads to even less diffusion than the high-resolution uniform scheme.  Comparing the distributions of
$\bsq$ between the two least diffusive runs (Fig.~\ref{fig:loop-evol-horizontal}-\ref{fig:loop-evol-cart-horizontal}), 
we see that less magnetic energy density has diffused from the center in the ``warped parabolic'' run than in the 
``uniform high-resolution parabolic'' run.   Even though we tuned the warped coordinates to be nearly uniform 
within the loop's region, it is not perfectly uniform:  there is $\sim 10\%$ variation in the grid spacing from the 
center of the loop to the loop's edge, making the grid spacing at the loop's center $\sim 5\%$ smaller
in the warped run than in the high-resolution uniform run.  This slight difference in grid spacing likely 
explains the differences between the two runs shown in Fig.~\ref{fig:bsq_diffusion} since they are of the 
same relative magnitude (of a few percent).  The fact that the warped run performs at least as well 
as the high-resolution uniform run verifies that the coordinate system was successfully implemented and 
is effective at moving a mesh refinement without introducing spurious behavior in the refined region.  

We conclude with a note that the parabolic method is the default 
choice of methods in \harm, and it is used in the rest of the paper. 


\section{Warped Spherical Coordinates}
\label{sec:warp-sph-coord}

Having obtained encouraging results with the warped Cartesian implementation, we have designed an analogous coordinate transformation adapted to spherical coordinates.
Recall that we are ultimately interested in evolving accretion disks around binary black holes, so our goal is a transformation that accurately resolves the region around the black holes, resolves the 
small aspect ratio of the disk, and conforms to the near azimuthally-symmetric shape of the disk past the binary's orbit. 
The full coordinate transformation will be a 3-d one, but let us describe the 2-d version since the transformation along the third dimension is straightforward.
We also outline first the (simpler) transformation where the two focal points occur at the same radial distance  from the coordinate origin, and then deal with the case 
where the focal points can be at different radii (e.g., for non-equal mass binaries).  In all cases, the center of 
mass of the binary will be fixed at the coordinate system's origin.

\subsection{Circular Azimuthal Focusing}
\label{sec:circ-azim-focus}

In this 2-d version of the warped coordinates, constructed to accommodate an equal-mass binary black hole spacetime, we consider only the transformation between $\left\{r,\phi\right\}$ and $\left\{x,y\right\}$, where $\left\{r,\phi\right\}$ are our physical coordinates, with $\phi$ being 
the periodic azimuthal coordinate and $r$ the radial coordinate. In the numerical $\left\{x,y\right\}$ coordinate system, $y$ will be the radial-like coordinate and $x$ will be the azimuthal-like one; we are therefore interested 
in keeping the periodicity assumption (Property~\ref{prop:dx-peridocity} from the previous section) along $x$, but not along $y$. Thus, for the $\phi$ coordinate, we can use the same transformation as before [Eq.~(\ref{X-eq-1})] (but without periodicity in $y$),
\begin{eqnarray}
\fl \phi(x,y)  = 2 \pi x 
 &{} - 2 \pi a_{x1} \ {\tau}_{y3}(y,t) \Big[ & \, \tilde{\mathcal{T}}_{x1}(x,t)  - \tilde{\mathcal{T}}_{x1}(x_1(t),t) \nonumber \\
  &{}&{} - \left( x - x_1(t) \right) \left( \tilde{\mathcal{T}}_{x1}(1,t) - \tilde{\mathcal{T}}_{x1}(0,t)  \right)   \, 
  \Big]   \nonumber \\
 &{} - 2 \pi a_{x2} \ {\tau}_{y4}(y,t) \Big[ & \, \tilde{\mathcal{T}}_{x2}(x,t)  
  -  \tilde{\mathcal{T}}_{x2}(x_2(t),t) \nonumber \\ 
 &{}&{} - \left( x - x_2(t) \right) \left( \tilde{\mathcal{T}}_{x2}(1,t)   - \tilde{\mathcal{T}}_{x2}(0,t)    \right)   \, 
  \Big]\label{phi-eq-warped}
\end{eqnarray}
where we have introduced the shorthand notations
\begin{eqnarray*}
\tilde{\mathcal{T}}_{xi}(x,t) & \equiv \tilde{\mathcal{T}}(x,x_i(t),\delta_{xi},h_{xi}) \quad , \\
{\tau}_{y3}(y,t) & \equiv {\tau}(y,y_1(t),\delta_{y3},h_{y3}) \quad , \\
{\tau}_{y4}(y,t) & \equiv {\tau}(y,y_2(t),\delta_{y4},h_{y4}) \quad . 
\end{eqnarray*}

\begin{figure}[tbhp]
\centering
\includegraphics[width=0.6\textwidth]{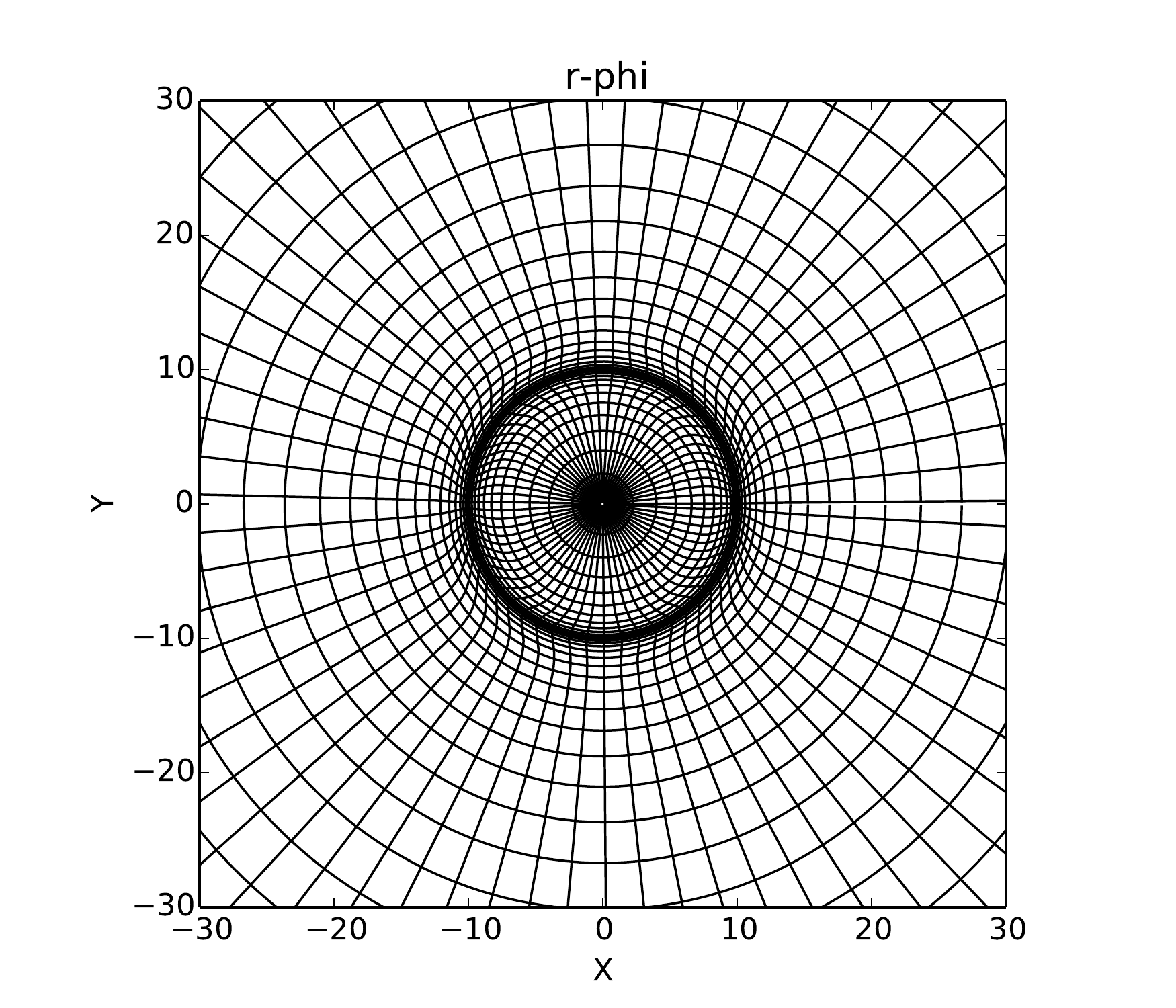}
\caption[]{Example of a warped spherical grid, using circular azimuthal warping only.  The parameters are 
$\delta_{x1}=\delta_{x2}=\delta_{y3}=\delta_{y4}=0.1$, 
$a_{x1}=a_{x2}=1$, $h_{x1}=h_{x2}=h_{y3}=h_{y4}=20$, 
$s=10^{-2}$, $b_{r}=6.5$, $\rin=0$, $\rout=300$, $x_1=0$, $x_2=0.5$, $r_{b} = 10$.
  \label{fig:sph-grid1} }
\end{figure}

For the $r$ coordinate, though, we are interested in a different transformation.
We want to focus cells in the vicinity of the black holes, but rarefy them at an exponential rate as one proceeds away from them---similar to what is done in single black hole 
accretion disk simulations (e.g.,~\cite{Noble09}). 
Rarefaction near the origin is motivated by the fact that the origin is not a point of special interest for our 
problem\footnote{As far as the authors can tell from previous work, no small scale phenomena develops at the 
center of mass of the system \cite{2010ApJ...715.1117B,Farris11,Bode12,2012PhRvL.109v1102F}.  This is to be expected
as the center of mass is a point of unstable equilibrium;  gas elements near it will adiabatically expand and 
fall toward the nearest black hole if left alone.}, and because there 
will already be many cells in its vicinity due to the focusing of azimuthal spacing that naturally arises there in spherical coordinates. 
Radial grid spacing is increased with radius in order to cover a larger radial extent and is justified because characteristic radial scales of the disk's 
turbulence tend to scale with their radius (at least for disks of constant scale height). 
We therefore consider the following relation
\begin{eqnarray}
 r(y) & = \rin + \left(b_{r} - s a_{r} \right) y + a_{r} \left[ \sinh\left(s(y-y_b)\right) + \sinh\left(s y_b \right) \right], \label{r-of-y-2} \\[0.3cm]
  a_{r} & \equiv \frac{ \rout - \rin - b_{r} }{ \sinh\left(s(1-y_b)\right) 
+ \sinh\left(s y_b\right)  - s} , \label{ar-const}
\end{eqnarray}
where $s$ controls the strength of the transition in resolution near $y_{b}$.
The meaning of the parameters is more readily gleaned 
when looking at $\partial r/\partial y$:
\begin{equation}
\pderiv{r}{y} = b_{r}  + s a_{r} \left[ \cosh\left( s (y-y_b) \right)  -  1  \right] . \label{dr-dy} 
\end{equation}
We see immediately that 
$b_{r}$ is the minimum of $\pderiv{r}{y}$ which occurs at $y=y_b$, 
$a_{r}$ is the amplitude of the nonlinear term, and that the radial grid spacing increases away from $y=y_b$. The radii $\rin$ and $\rout$ are
defined as the radius of the inner edge of the domain and the radius of the outer edge of the domain, respectively. It is implicit in this transformation that $\yin = 0$ and $\yout = 1$, because of Property~\ref{prop:xy-XY-bounds}
of Section~\ref{sec:implementation}. The parameter $y_b$ is determined by inverting
$r(y_b) = r_b$ numerically using~(\ref{r-of-y-2})\footnote{The inversion procedure is performed via a Newton-Raphson scheme.} where $r_b$ is the radius of the black holes' orbit. The grid is therefore
parameterized by $\{b_r, \rin, r_b, \rout, s\}$. In Figure~\ref{fig:sph-grid1} we show an example.

\subsection{General Warped Coordinates}
\label{sec:bin-warp-coord}

As defined in the previous section, $r(y)$ will have one focal point at $r=r_b$ for all
$\phi$. When evolving black holes with unequal mass ratios in the center of mass frame, however, the
radial coordinates of the black holes will be different.  
In this section we will generalize the transformation from the previous section to accommodate focusing at different radii as well as introduce the dependence along the third dimension.

In order to focus resolution at different radii, we will smoothly interpolate
between different $r(y)$ that use three different sets of $\{b_r, r_b\}$, two
for the black holes and one for the intermediate region:
\begin{equation}
r(x,y) = r_3 + \left(r_1 - r_3\right) \tilde{\tau}(x,x_1,\delta_{x 1}) + \left(r_2 - r_3\right) \tilde{\tau}(x,x_2,\delta_{x 2}) , \label{V-gen-interp-func}
\end{equation}
where $r_{1,2,3}=r_{1,2,3}(y)$ [given by Eq.~(\ref{r-of-y-2})], $r_{1,2}$ have values tailored to the black hole at---respectively---$x_{1,2}$.  The intermediate 
profile,  $r_3(y)$,  is the value of $r(y)$ at the azimuthal points intermediate to the two black holes. 
Hence, the numerical inversion of $r(y_{i}) = r_{bi}$ ($i=1,2,3$)
needs to be calculated for all $x$ and whenever the black hole positions change.
We set $b_{1,2,3}$ (we omit the subscript $r$ in the parameters $b$ from now on to simplify the notation) depending on the resolution requirements.  We must set
$b_{3} > b_{1,2}$ to focus more radial zones at $\phi_{1,2}$. 
In practice, we see no reason to have $\rin$ and $\rout$ different between the
regions, as this ensures that the inner and outer edges of the domain are at
constant radii which better accommodates the outflow boundary conditions imposed there.
The circular azimuthal focusing coordinate system described in Section~\ref{sec:circ-azim-focus} is
a special case of the general warped system when $r_1=r_2=r_3$.

\begin{figure}[tbp]
\centering
\includegraphics[width=0.6\textwidth]{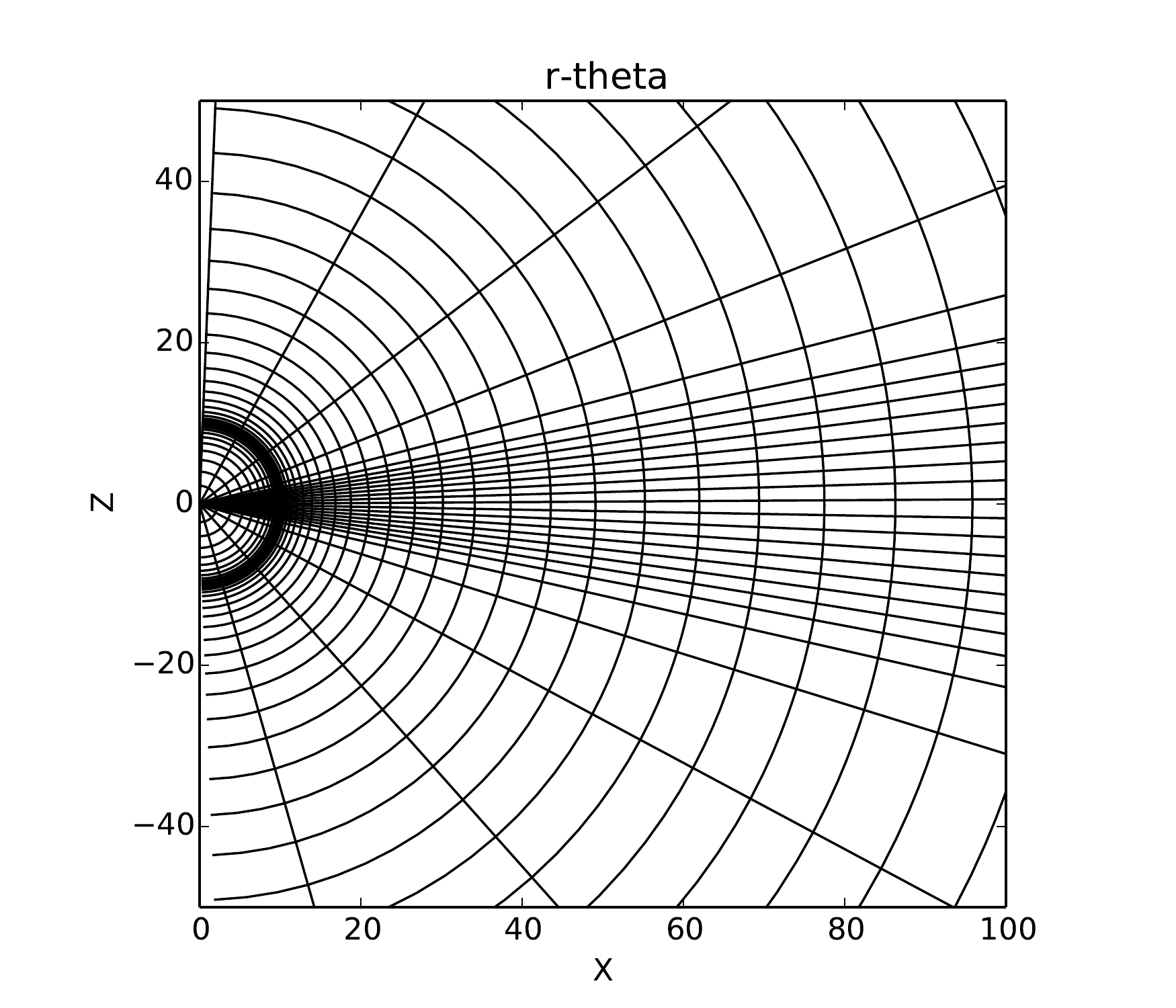}
\caption[]{Poloidal slice of a warped spherical grid, with parameters 
$\delta_{x1}=\delta_{x2}=\delta_{y3}=\delta_{y4}=0.1$, $\delta_z = 0.4$, 
$a_{x1}=a_{x2}=1$, $a_z = 4$, 
$h_{x1}=h_{x2}=h_{y3}=h_{y4}=h_{z}=20$, 
$s_i=10^{-2}$, $b_i=6.5$, $\rin=0$, $\rout=300$, $x_1=0$, $x_2=0.5$, $r_{b1} = r_{b2} = r_{b3} = 10$.  The equatorial slice of these 
coordinates is identical to that shown in Figure~\ref{fig:sph-grid1}. \label{fig:sph-grid2} }
\end{figure}
\begin{figure}[tbp]
\centering
\includegraphics[width=0.45\textwidth]{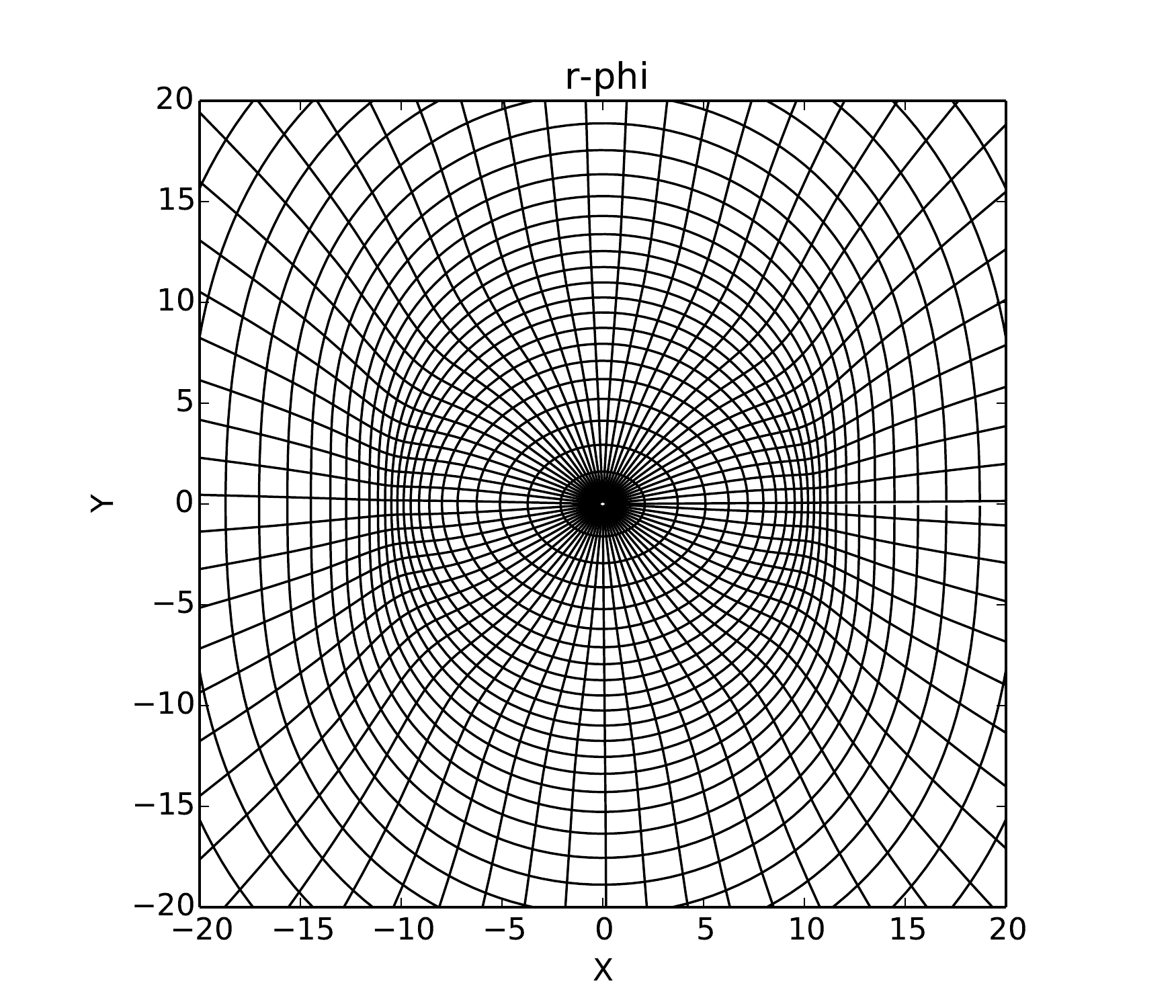}
\includegraphics[width=0.45\textwidth]{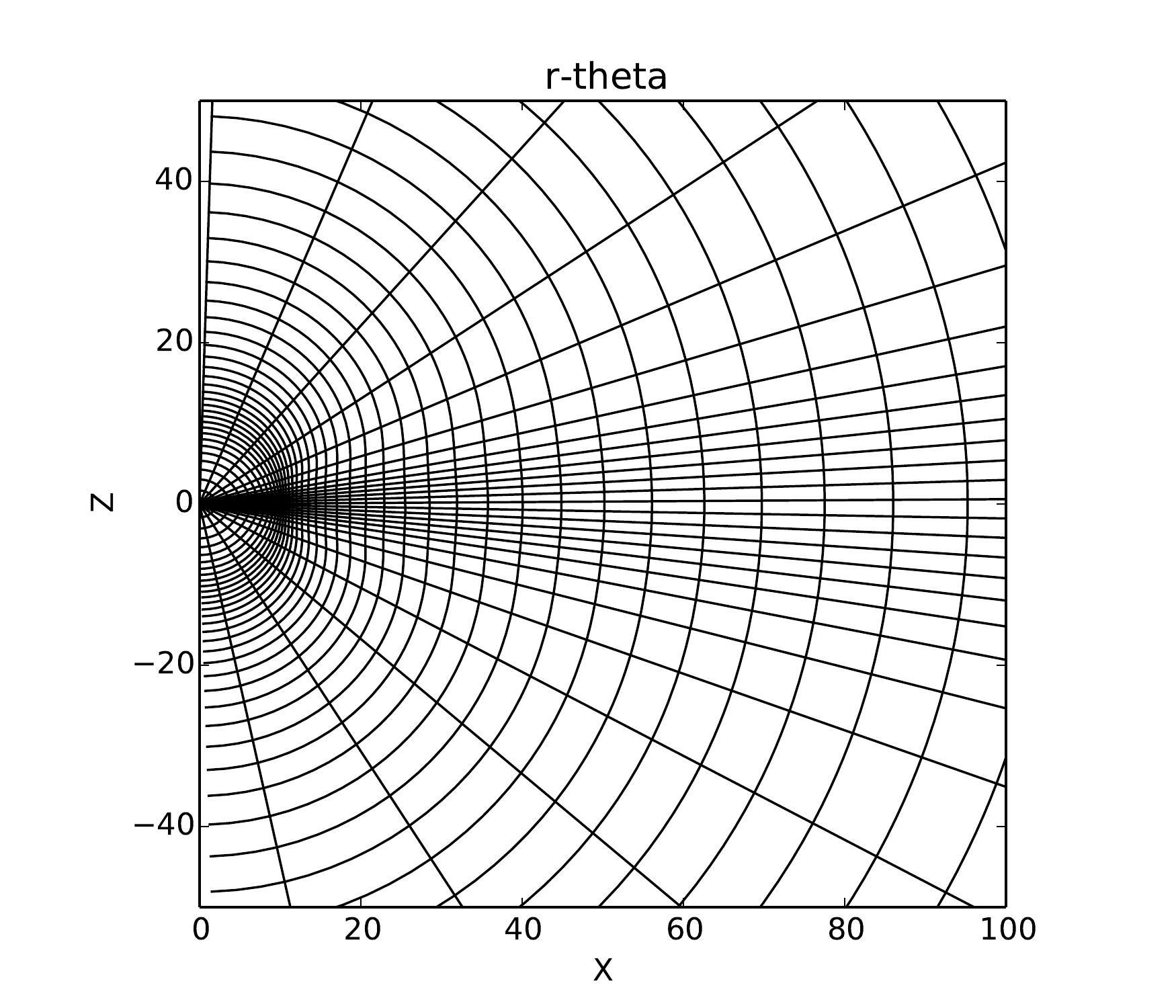}
\caption[]{Poloidal and equatorial slices of a different warped spherical grid configuration, with parameters $\delta=0.1$, $\delta_z = 0.4$, $a_{x1}=a_{x2}=1$, $a_z = 4$, $h_{x1}=h_{x2}=h_{y3}=h_{y4}=h_{z}=10$, $s_{1,2}=10^{-2}$, $s_{3}=4$, $b_{1,2}=6.5$, $b_{3}=40$, $\rin=0$, $\rout=300$, $x_1=0$, $x_2=0.5$, $r_{b1} = 10$, $r_{b2}=15$, $r_{b3}=12$.  
\label{fig:sph-grid3} }
\end{figure}


The previous set of coordinates lacks dependence on poloidal angle ($\theta$). We introduce this dependence here. 
Like $x$ and $y$, we will assume that $z \in \left[0,1\right]$.

Like that used in previous calculations, the poloidal coordinate is refined near
the equator.  We need not distort it further about each black hole as the
resolution in the midplane is already sufficiently fine.
Using a construction similar to that already presented
\begin{equation}
\theta(z) = \pi \left\{  z - a_z  \left[ \tilde{\mathcal{T}}_{z1}(z) - \tilde{\mathcal{T}}_{z1}(z_1) 
- \left(z-z_1\right) \left( \tilde{\mathcal{T}}_{z1}(1) - \tilde{\mathcal{T}}_{z1}(0) \right) \right] \right\}
\label{eq:theta-warped-final}
\end{equation}
where we will fix the $z$ coordinate of the focus to be $z_1 = 1 / 2$ and not a function of time like $x_1, x_2, y_1, y_2$ (though in principle it could be), since the black holes are always located in the equatorial plane.


Far from the black holes, we want the coordinates to approach uniform spherical coordinates.  
This is done by multiplying the distortion terms by $z$-dependent boxcar functions: 
\beqa{
\fl r(x,y,z,t) = r_3(y,t) + \tilde{\tau}_{z1}(z)
 \Big\{ & \left[r_1(y,t) - r_3(y,t)\right] \tilde{\tau}_{x3}(x,t) \nonumber \\
    & \quad + \left[r_2(y,t) - r_3(y,t)\right] \tilde{\tau}_{x4}(x,t) 
\Big\} , \label{r-warped-final} 
}
\beqa{
\fl \phi(x,y,z,t) = 2 \pi \Bigg\{  x  - \tilde{\tau}_{z1}(z) & \bigg[ 
 a_{x1} \ {\tau}_{y3}(y,t) & \Big( \, \tilde{\mathcal{T}}_{x1}(x,t)  - \tilde{\mathcal{T}}_{x1}(x_1(t),t) \nonumber \\
&{}&{}  - \left( x - x_1(t) \right) \left( \tilde{\mathcal{T}}_{x1}(1,t) - \tilde{\mathcal{T}}_{x1}(0,t)  \right)  \Big)  \nonumber  \\
  & + a_{x2} \ {\tau}_{y4}(y,t) & \Big( \, \tilde{\mathcal{T}}_{x2}(x,t)  
   -  \tilde{\mathcal{T}}_{x2}(x_2(t),t) \nonumber \\
 &{}&{} - \left( x - x_2(t) \right) \left( \tilde{\mathcal{T}}_{x2}(1,t)   - \tilde{\mathcal{T}}_{x2}(0,t)    \right) \Big) \bigg] \! \Bigg\}
 \label{phi-warped-final}
}
where
\begin{equation}
\fl r_i(y,t) = \rin + b_{i} y  + \left(\rout - \rin - b_{i} \right)
\left[ \frac{ \sinh\left[s_{i}(y-y_{i}(t))\right] + \sinh\left[ s_i y_{i}(t) \right] - s_i y }{
 \sinh\left[s_i(1-y_{i}(t))\right] + \sinh\left[ s_i y_{i}(t)\right] -s_i  }\right]
\label{ri-eq-2}
\end{equation}
and, as in the previous section,
\begin{eqnarray*}
{\tau}_{y3}(y,t) & \equiv {\tau}(y,y_1(t),\delta_{y3},h_{y3}) \\
{\tau}_{y4}(y,t) & \equiv {\tau}(y,y_2(t),\delta_{y4},h_{y4}) \\
\tilde{\tau}_{x3}(x,t) & \equiv  \tilde{\tau}(x,x_1(t),\delta_{x3},h_{x3}) \\
\tilde{\tau}_{x4}(x,t) & \equiv  \tilde{\tau}(x,x_2(t),\delta_{x4},h_{x4}) \\
\tilde{\tau}_{z1}(x,t) & \equiv  \tilde{\tau}(z,z_1(t),\delta_{z},h_{z}) \\
\tilde{\mathcal{T}}_{x1}(x,t) & \equiv  \tilde{\mathcal{T}}(x,x_1(t),\delta_{x1},h_{x1}) \\
\tilde{\mathcal{T}}_{x2}(x,t) & \equiv  \tilde{\mathcal{T}}(x,x_2(t),\delta_{x2},h_{x2}) \\
\end{eqnarray*}
We thus arrive at the final form of our coordinate transformation, comprised of equations~(\ref{eq:theta-warped-final},\ref{r-warped-final},\ref{phi-warped-final}).
Figures~\ref{fig:sph-grid2} and \ref{fig:sph-grid3} showcase possible grid configurations.
The final grid has a number of free parameters which need to be manually set on a case-by-case basis depending on the physical situation at play. In the following we try to summarize the meaning of these parameters.
\begin{itemize}
\item $\rin$ and $\rout$ are the radial coordinates of the inner and outer edges of the physical domain;
\item $x_1, x_2, y_1, y_2$ are the $(x,y)$ coordinates of the two focal points (black holes). $y_3$ is the $y$ coordinate of the intermediate region focal point. In general these have to be set at each time-step, and for our specific implementation this is done as follows: let $r_1, r_2, \phi_1, \phi_2$ be the $(r,\phi)$ coordinates of each black hole at every time-step (which are known); $r_3 = \max(r_1, r_2)$, we set $x_{1,2}$ to be 
\begin{equation}
\label{eq:x12-of-phi12}
x_{1,2} = \frac{\phi_{1,2} \bmod{2\pi}}{2\pi}
\end{equation}
and we compute $y_{1,2,3}$ via a Newton-Raphson root-finding procedure\footnote{We note that one would in principle have to invert relations~(\ref{r-warped-final}) and (\ref{phi-warped-final}), but we find that (\ref{eq:x12-of-phi12}) and (\ref{eq:y123-of-r123}) give sufficient accuracy for our purposes.}
\begin{equation}
\label{eq:y123-of-r123}
r_{1,2,3}  \left(y_{1,2,3}\right) = r_{1,2,3} \, .
\end{equation}
$z_1$, the $z$ coordinate of the focal points, is kept fixed at $z_{1} = 1/2$ (the equatorial plane).
\item $\delta$ and $h$ parameters are related to the width and steepness of the approximate (non-ideal) step-functions, which control the extent of and transition rate to the warped regions.
\item $b_i$ is the minimum of $\pderiv{r_i}{y}$ which occurs at $y=y_i$, and the transition steepness is controlled by $s_i$.
%
\item $a_{x1}, a_{x2}, a_{z}$ are coupling parameters. 
\end{itemize}

\begin{figure}[tbhp]
\centering
\includegraphics[width=0.45\textwidth]{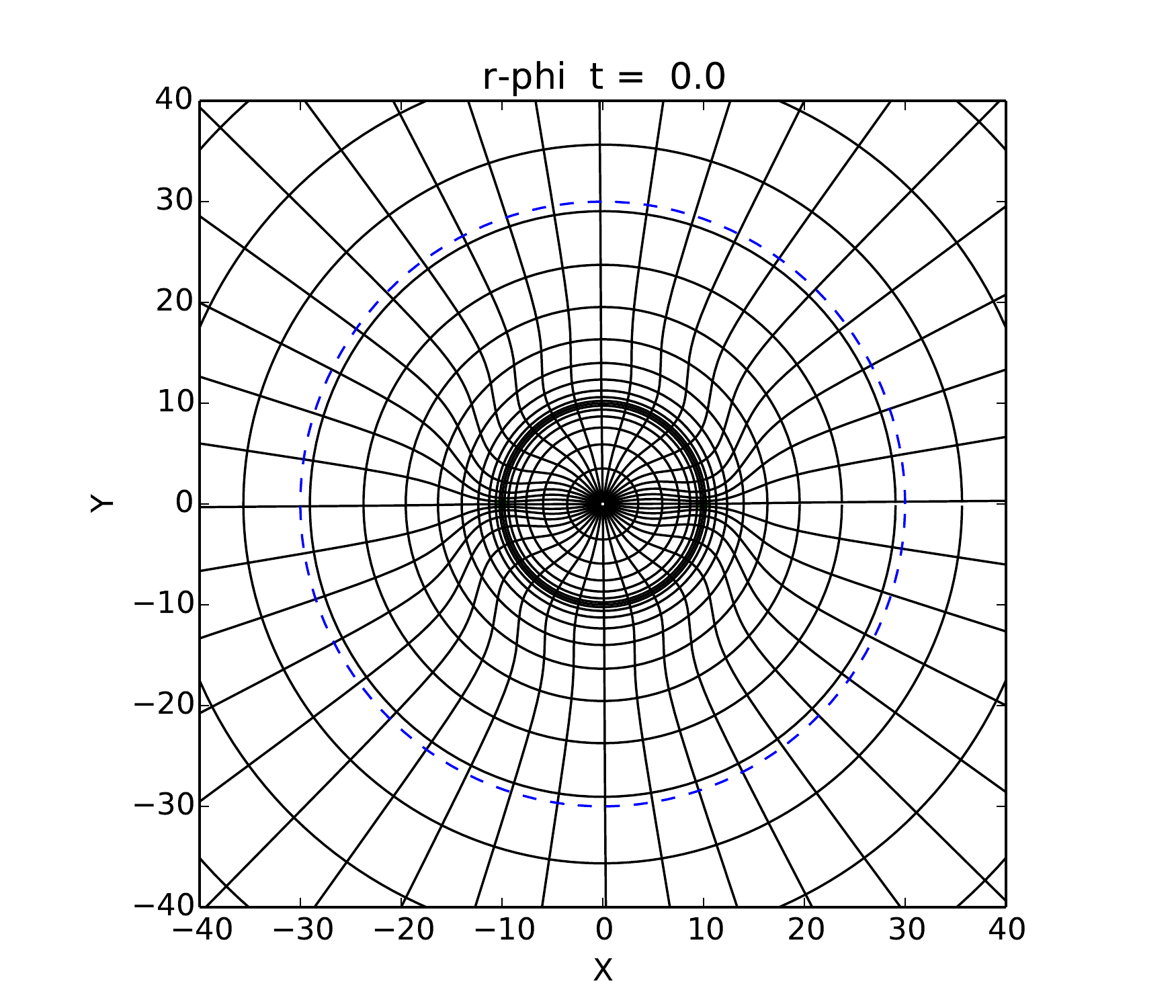} 
\includegraphics[width=0.45\textwidth]{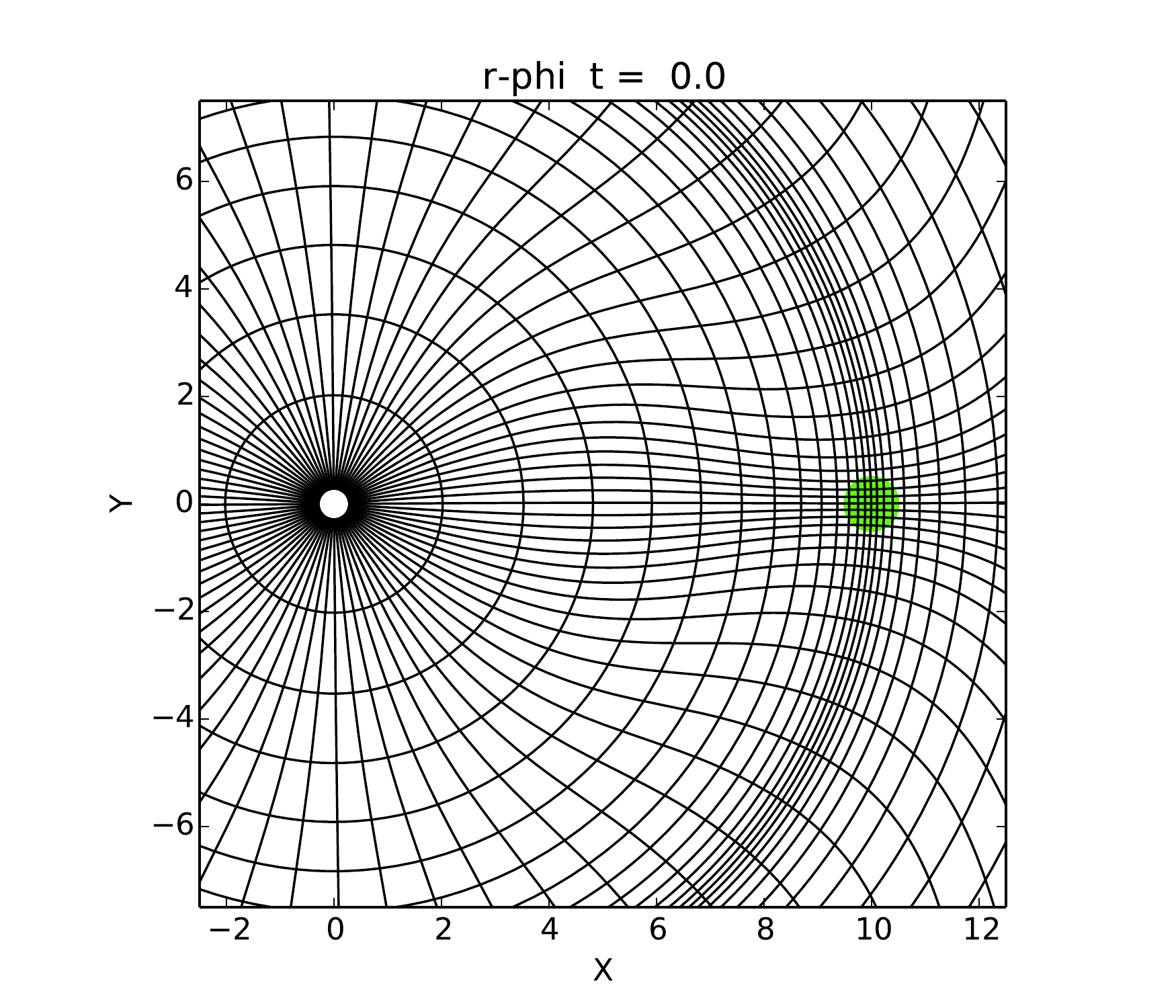} \\
\includegraphics[width=0.45\textwidth]{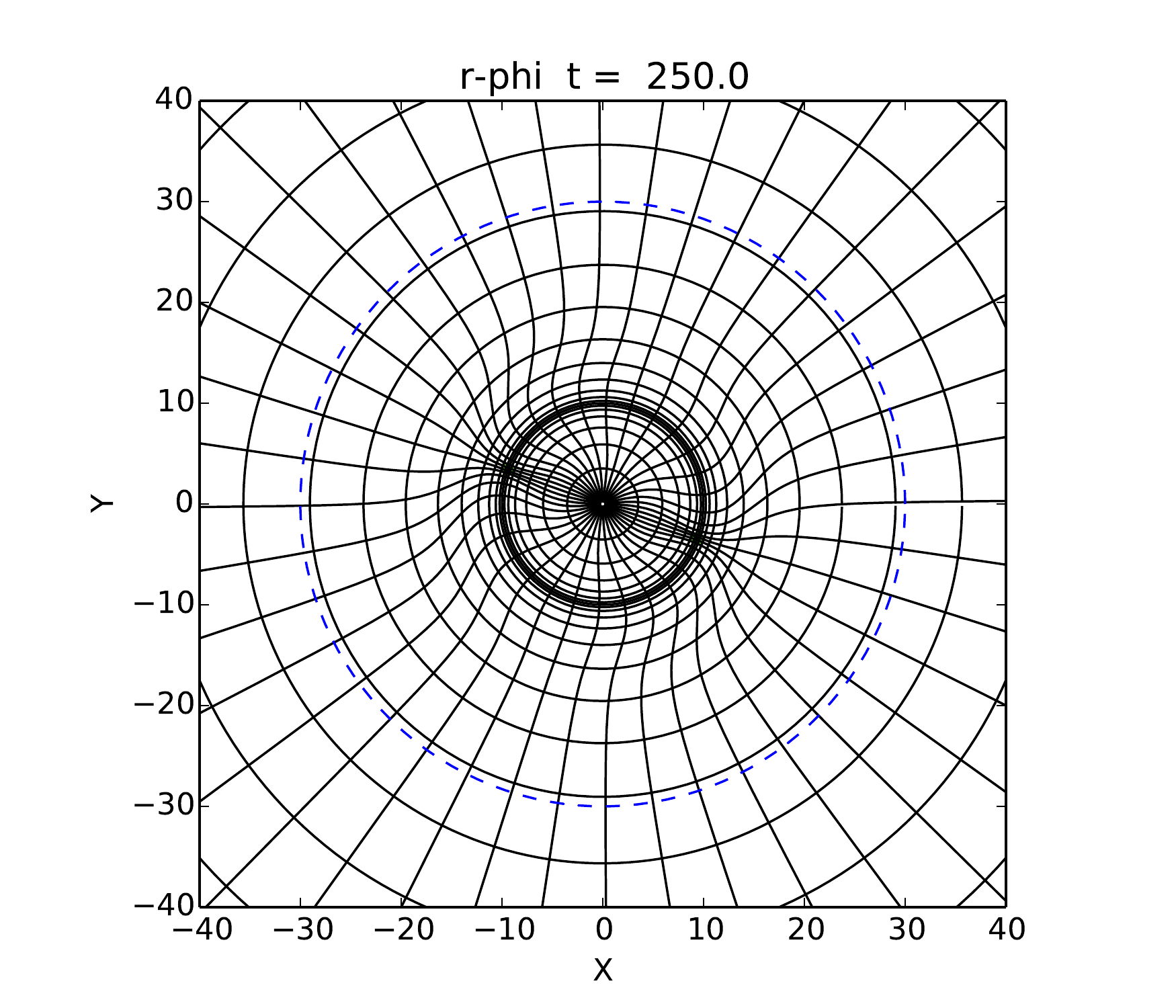} 
\includegraphics[width=0.45\textwidth]{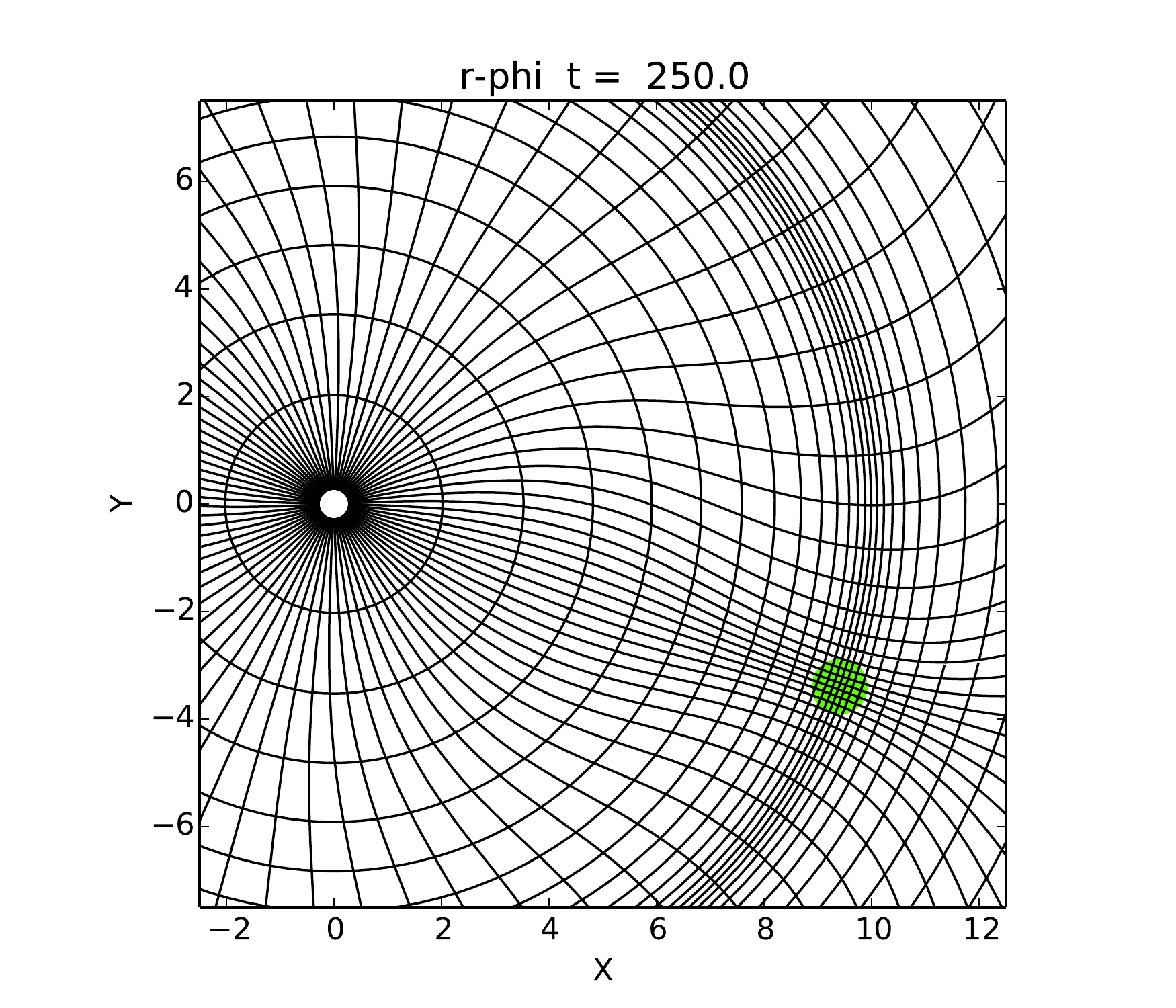} \\
\includegraphics[width=0.5\textwidth]{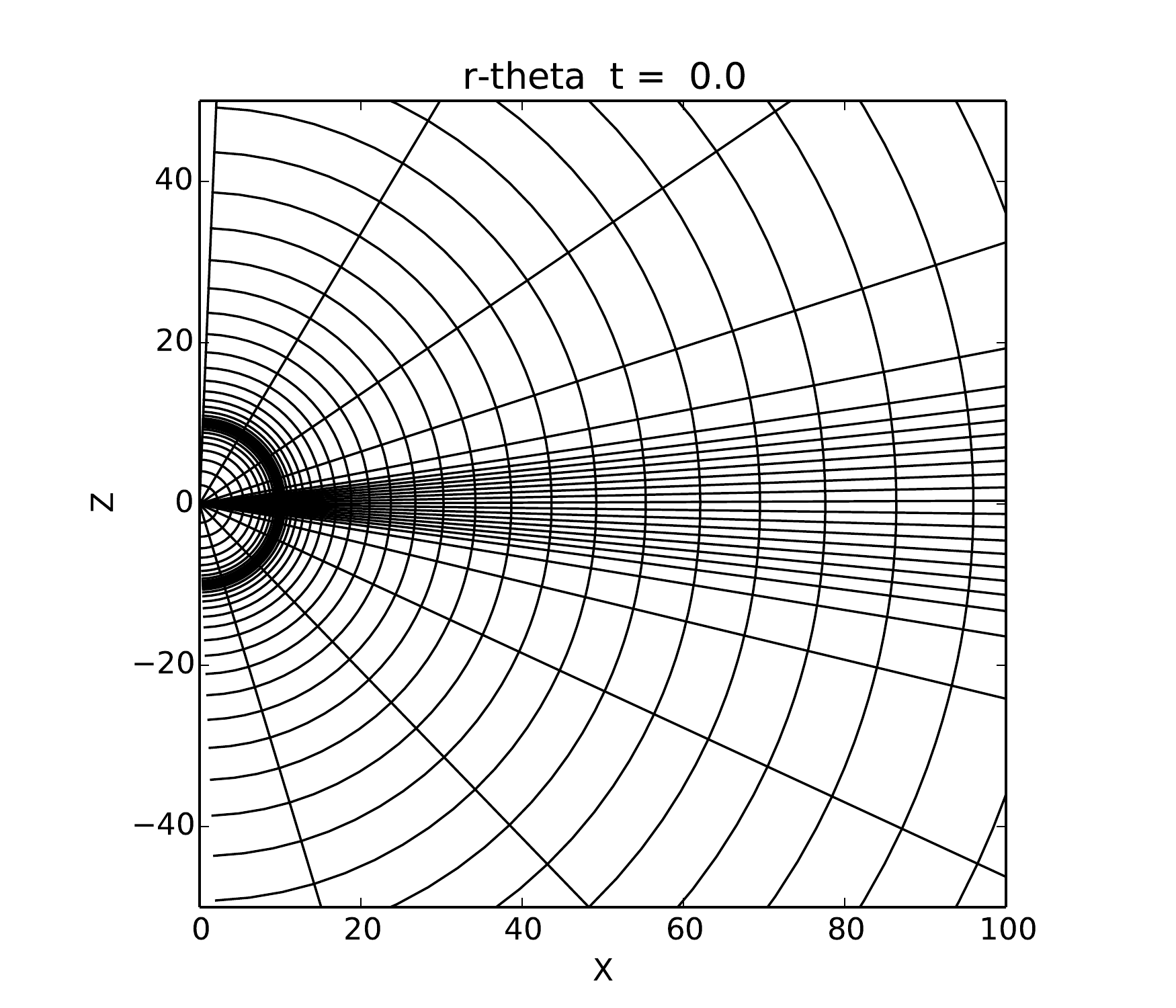}
\caption[]{Example grid used for a binary black hole system. The green-shaded area marks the black holes' event horizons, and the dashed blue line marks the location of the accretion disk's inner radius. 
Grid parameters are as follows:
$\delta_{x1}=\delta_{x2}= 0.1$, $\delta_{x3}=\delta_{x4}=\delta_{y3}=\delta_{y4}=0.18$, 
$a_{x1}=a_{x2}=1.5$, $a_{z}=4.3$, $h_{x1}=h_{x2}=h_{x3}=h_{x4}=h_{y3}=h_{y4}=h_z=20$, 
$s_{i}=10^{-2}$, $b_{i}=6.5$, $\rin=0$, $\rout=300$, $r_{b} = 10$; $x_1$ and $x_2$ are set at each time-step to coincide with the location of the black holes. Parameters are chosen so that there are approximately $32$ cells spanning each black hole horizon in each dimension (black hole horizons span $M \equiv M_{\rm BH1}+M_{\rm BH 2} = 1$ in this coordinate system). 
\label{fig:bbh-grid}  }
\end{figure}

Figure~\ref{fig:bbh-grid} depicts the grid configuration we devised for our evolutions of a binary black hole system. The grid is dynamic, with the location of the focal points tied to the location of the binary. 

\subsection{Implementation Details and Computational Effort}

In terms of computational cost, evaluating $x^\nu(\xp{\mu})$ and their derivatives as they are written here for each cell would be computationally intensive and account for a large 
fraction of the runtime, largely because they involve expensive transcendental functions (e.g., $\tanh$).   
One way we have dramatically reduced the computational cost is by separately computing functions by their dependent 
variables---Eqs.~(\ref{eq:theta-warped-final},\ref{r-warped-final},\ref{phi-warped-final}) can each be expressed
as a sum of products of functions that each depend only on time and one of the $\xp{i}$ variables.  These time-dependent 1-d functions (and their derivatives) are evaluated once per time step along 
each dimension and stored in 1-d arrays, which are used in all the necessary coordinate calculations thereafter.  Since our simulations are 3-d, several 1-d function evaluations amount to a trivial fraction of runtime.  

That is not to say that the overhead of the coordinate transformation is insignificant;  the effort it takes transforming all tensors to and from our dynamic 
coordinates---i.e.\ a la Eq.~(\ref{general-coordinate-transformation})---is larger than any other operation related to the dynamic coordinate system 
and amounts to be about $\simeq 6-11\%$ of runtime for a typical circumbinary disk evolution.   However, these operations are necessary in our simulations anyway because our spacetime metric is
defined in Cartesian coordinates and we choose to simulate the gas in spherical coordinates~\cite{Noble:2012xz}.   As a result, our circumbinary disk runs suffer very little additional effort 
($\lesssim 1-2\%$ of runtime) from using dynamic coordinates instead of static coordinates, at least for the 
binary black hole runs that we will describe later in the paper.  Since the computations
made for the dynamic coordinates are performed locally on a processor per cell per time step, 
their contribution to the overall effort does not depend on the number of processors used in the run and should not 
depend strongly on the computer's architecture.  Our timing measurements were made using 
the Stampede cluster at the Texas Advanced Computing Center and the BlueSky cluster at RIT. 


\section{Results}
\label{sec:results}

To test our implementation and measure the effect of the warped coordinate system, we have performed a number of tests and, when possible, matched the results against those obtained using ``standard'' well-tested coordinate 
systems present in the \harm code. 
We present below results obtained for the evolution of: the magnetized Bondi solution, Section~\ref{sec:bondi}; an accretion disk in the background of a single black hole, Section~\ref{sec:disk-singleBH}; a circumbinary accretion disk, Section~\ref{sec:disk-BBH}. 

\subsection{Bondi Flow}
\label{sec:bondi}

The Bondi solution is a solution to the equations describing  spherically symmetric, time-independent accretion of non-magnetized gas onto a central, perfectly absorbing object (e.g., a black hole)~\cite{Bondi:1952ni}.
The particular solution we use here has $\dot{M} = 4\pi r^2 \rho u^r = 4\pi$, the adiabatic index of the equation of state is $\gamma = 4/3$, and the radial coordinate of the sonic point is $r_s = 8M$. 
Our line element is the Schwarzschild solution written in Kerr-Schild coordinates.  We add a weak, radial magnetic field to the initial conditions in order to validate the magnetic field evolution routines, too.
Adding a radial magnetic field to the system does not alter the solution analytically, but it does test how a code handles 
terms in the EOM at the truncation error level that do not satisfy the Bondi equations.   Relative to evolutions in standard spherical coordinates, we expect these terms to result in a larger effect 
for our warped coordinates as the system no longer conforms to the symmetry of the problem. 

For this test, we use the warped spherical system described in Section~\ref{sec:bin-warp-coord} using a similar setup to that illustrated in Figure~\ref{fig:sph-grid1}.
We stress that this grid was \emph{not} tailored for this example; our goal here is not to evolve the Bondi solution as accurately as possible, but rather to measure possible 
artifacts introduced by the warped and dynamic grid during the evolution.  We compare the warped evolution to a 
run  with a coordinate configuration we call ``unwarped,'' which is the same as the warped coordinate setup 
except we move the focal points to the origin by setting $r_{b1} = r_{b2} = r_{b3} = 0$ and $a_{x1}=a_{x2}=0$.   
The unwarped setup is static, azimuthally symmetric, and is similar to the grids used in single black hole accretion 
disk studies where $\Delta r/r$ is constant~\cite{Noble09}. 

We follow the procedure outlined in~\cite{GMT03} to construct the initial data. We have performed pure hydrodynamic as well as magnetized evolutions. For the magnetized cases, we prescribe a vector potential with the form
\begin{equation}
\label{eq:vec-pot}
A_{\mu} = \left(0, 0, 0, -A_0 \cos \theta \right) \,,
\end{equation}
which guarantees a radial divergence-less magnetic field with
\begin{equation}
\label{eq:bsq-bondi}
\bsq = \frac{A_0^2}{r^4} \,.
\end{equation}

\begin{figure}[htbp]
\centering
\includegraphics[width=0.8\textwidth]{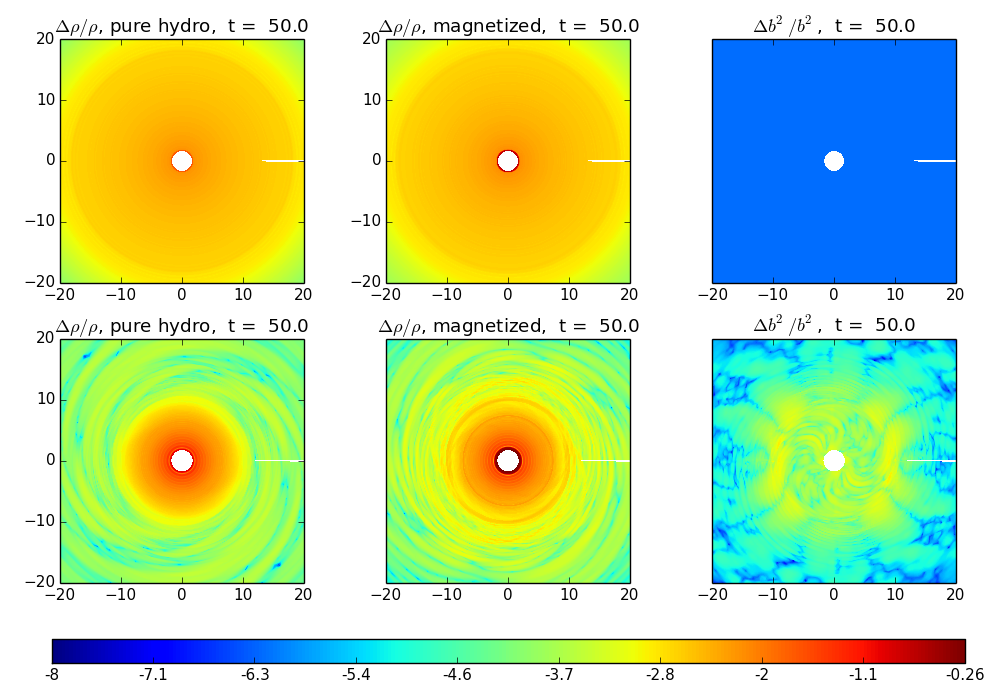}
\caption[]{Contour plots (in logarithmic units) of the relative change in
  density ($\rho$) [Eq.~(\ref{eq:bondi-rel-change}), left and middle columns]
  as well as the relative error in magnetic energy density ($\bsq$)
  [Eq.~(\ref{eq:magbondi-rel-change}), right column] at $t=50M$ for a 2-d
  equatorial Bondi solution test.  The pure hydrodynamic Bondi test (left column) 
  is compared with the magnetized case (middle and right columns).
  Evolutions were made with the warped coordinates (bottom row), and for an
  unwarped coordinate system with the distortions effectively moved to the
  origin (top row); see text for the parameters we used.  The narrow white wedge
  appearing in all plots is a visualization artifact that marks the edge between
  the beginning and end of $\xp{3}$.  
  \label{fig:bondi-rel-change} }
\end{figure}

The problem was integrated until $t = 50M$ in a 2-d equatorial domain, with $r \in [1.7, 100]M$, $256 \times 256$ cells and a time step of $dt = 0.01~M$.
We used a warped spherical grid with all the parameters set the same as in Figure~\ref{fig:sph-grid1}, except here we use $\rin=1.7$ and $\rout=100$ instead.  
We fixed the orbital angular velocity of the (fictitious)  black holes---or the grid's focal points---to be $\omega = 2\pi/50 $. 
The orbital frequency of the warp used here is more than $10$ times as large as what would be used for a binary evolution since the binary's 
orbital period is $P_\mathrm{orb} \simeq 561 M$ at this separation ($20~M$). 
Since Bondi is a steady-state solution, we (ideally) expect to see no evolution in our primitive variables. Since the grid \emph{does} evolve, however, in order to quantitatively measure deviations from a variable's initial state \emph{at the same physical points}, we measure such deviations after exactly one orbital period:
\begin{equation}
\label{eq:bondi-rel-change}
\frac{\Delta \rho}{\rho} = \left| 
\frac{\rho(t=50M) - \rho(t=0)}{\rho(t=0)}
\right| \,.
\end{equation}
For our specified angular velocity of $\omega = 2\pi/50 $, this guarantees that we are performing the comparison at the same physical points. Figure~\ref{fig:bondi-rel-change} shows a contour plot of this quantity; inspecting the figure, we see relative differences of the order of $10\%$ close to the black hole horizon for the pure hydrodynamic case, and values close to $50\%$ for the magnetized case, which may at first seem worrisome. 
These rather large values can be explained by the fact that, as we have already stressed, the grid we used for these evolutions was tailored for binary black hole evolutions, \emph{not} for single black hole ones. Thus, by construction, the radial grid spacing increases as one approaches the black hole horizon (as can be observed in Figure~\ref{fig:sph-grid1}, for example, where the black holes would be located at $r=10~M$ with horizon radii of $0.5~M$), resulting in fewer cells there than in the unwarped setup shown here. 
This is particularly critical in the magnetized case, where the larger values of the magnetic field close to the black hole horizon would require a larger number of cells to keep the truncation error constant. 
We also note that the relative solution errors seem to equilibrate quickly, with very little evolution happening after $t \sim 7M $.
By increasing the overall number of cells we have further verified that these relative changes do decrease, and thus obtained further evidence that the relatively large values observed are indeed artifacts of insufficient resolution in the corresponding region, rather than an error or problem with our implementation.

For the magnetized case, a simpler quantitative test is achieved by simply evaluating the following quantity throughout the evolution
\begin{equation}
\label{eq:magbondi-rel-change}
\frac{\Delta \bsq}{\bsq} = \left| \frac{ \bsq r^4 - A_{0}{}^2 }{A_0{}^2}
\right| \,.
\end{equation}
We plot this quantity at $t=50M$ in Figure~\ref{fig:bondi-rel-change}, which demonstrates much less deviation in $\bsq$ from its initial condition than what was seen in $\rho$.   Apparently, our evolution 
procedure (including, e.g., our parabolic FluxCT scheme) introduces an insignificant level of numerical error into the magnetic field when using warped coordinates.

As a point of reference, we have measured the relative errors in $\rho$ and $\bsq$ for the magnetized Bondi test when using unwarped coordinates, which were described previously. The differences we find between the two 
sets of figures then highlight the effects from the truncation errors introduced by the warped coordinates.  
As expected for the unwarped case, we see no variation in relative errors in the azimuthal dimension.  The significantly lower 
magnetic field errors in the unwarped system are likely because only one of the induction equations is nontrivial (e.g., $\partial_t B^1 \ldots$), whereas---in the warped case---two 
induction equations are nontrivial which allows the magnetic field components to dynamically interact with each other. 
Also, the largest values observed with the warped coordinates occur in the immediate vicinity of the black hole horizon, where resolution is coarser than in the unwarped case; 
the unwarped system's radial grid spacing shrinks with diminishing radius, unlike the warped system whose radial grid spacing continuously grows from $r=10~M$ to the horizon.
Specifically, the radial step size in the vicinity of the black hole horizon is roughly $dr \simeq 0.056M $ in the unwarped case, as opposed to $dr \simeq 0.27M $ for the warped case.
Even so, the contrast between the density errors is not large (i.e.\ it is within a factor of 2) near the horizon, and the warped system becomes more accurate at larger radii since its radial resolution 
there is higher than that of the unwarped system.

\subsection{Disk with single black hole}
\label{sec:disk-singleBH}

In order to test the effect of the coordinates' distortion on an accretion disk, we have evolved an accretion disk in the background of a single (non-spinning) black hole with a numerical grid adapted to the binary configuration of Figure~\ref{fig:bbh-grid}. 
For details on the initial data and evolution procedure we refer the reader to~\cite{Noble:2012xz}.

We performed a 2-d equatorial evolution, with $400 \times 400$ cells;
the disk was initialized with an aspect ratio of $H/r=0.1$\footnote{Since this setup was assumed to be independent of polar angle, the aspect ratio serves as a 
parameter to control the pressure and angular velocity distribution in the disk and not the geometrical shape of the disk.}, inner edge at $r_{\rm in} = 40M$, and pressure maximum at $r_p = 78M$. 
Our motivation behind our choice of the disk's location and extent was to approximate the quasi-steady state of the circumbinary disk we have seen develop about an equal mass black hole binary~\cite{Shi:2011us,Noble:2012xz}. 
We evolved the configuration until $t=16000M$, corresponding to roughly 10 orbits at the inner edge of the disk, $3.7$ orbits at its pressure maximum, and about $28$ orbits of the warped grid's focal points. 

\begin{figure}[thbp]
\centering
\includegraphics[width=0.7\textwidth]{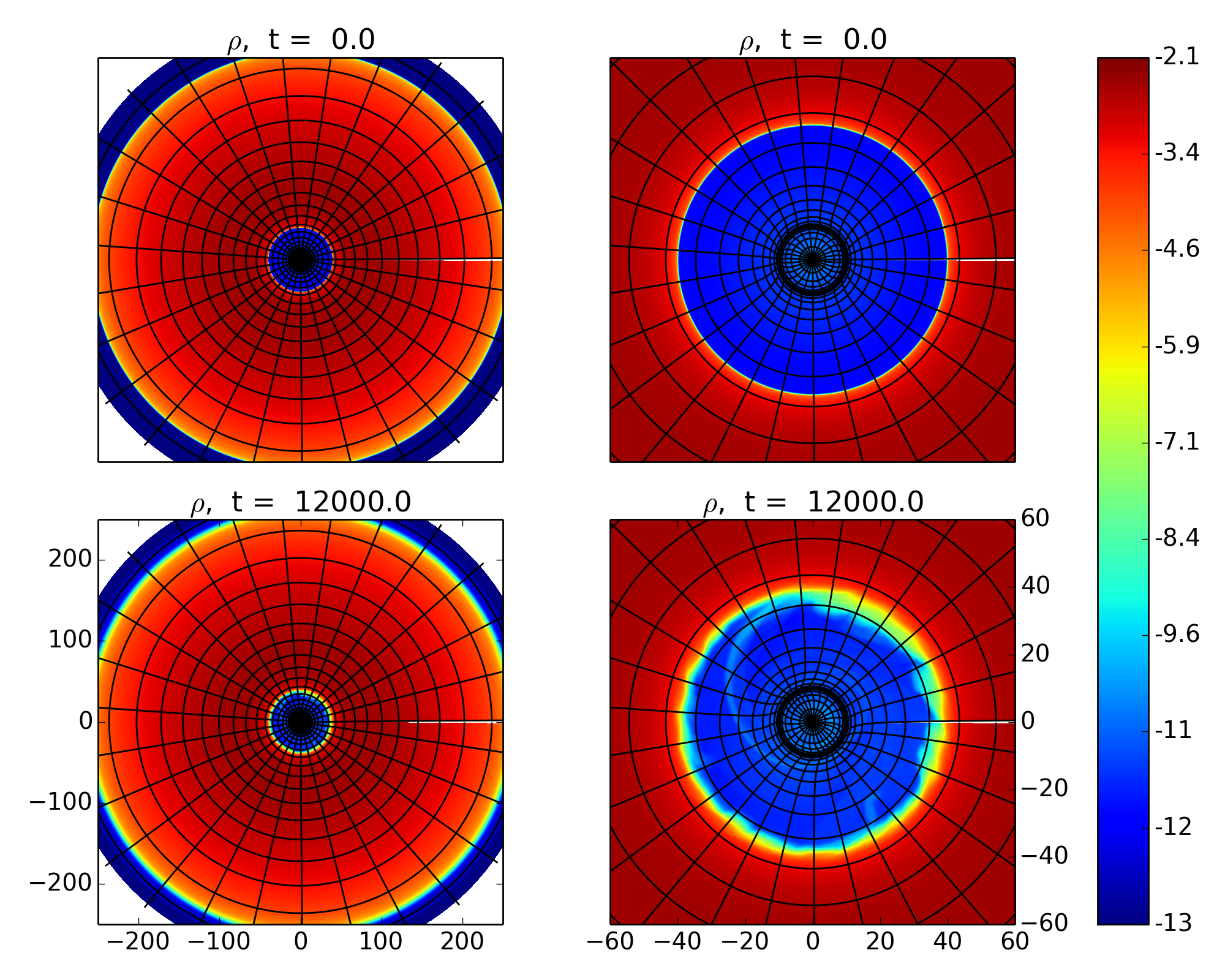}
\caption[]{Snapshots of the fluid density $\rho$ (logarithmic scale) at $t=0$ and $t=12000M$ for an accretion disk evolution in the background of a non-spinning single black hole. We have used the warped spherical grid of Figure~\ref{fig:bbh-grid}. 
The narrow white wedge appearing in all plots is a visualization artifact that marks the edge between the beginning and end of $\xp{3}$. 
 \label{fig:log10rho-warped} }
\end{figure}

Snapshots of the fluid density $\rho$ for two different time steps can be seen in Figure~\ref{fig:log10rho-warped}.
Again, we would like to emphasize that the black hole at the origin is not as well resolved as it typically is in single BH accretion disk simulations.  As a result, there are fluctuations in the fluid---that begin 
with density and pressure amplitudes close to that of the atmosphere---that propagate from near the horizon to the inner edge of the disk.  The disk is further perturbed by the time-varying azimuthal spacing 
of the warped coordinate system.  By $t \sim 12000M$, we can see perturbations have grown near the inner edge of the disk to amplitudes of a few orders of magnitude above the atmosphere level, which causes matter to start inflowing. 
To give a sense of scale, it takes an equal mass binary at the separation of the warps ($20~M$) approximately $13000M$ of time to inspiral.  In addition, the outer edge of the disk appears to remain unperturbed after its initial 
relaxation to a shallower gradient.  
We expect that magnetic stresses will dramatically alter this picture and 
give rise to a much more rapid development of accretion onto the black hole(s).  Our neglect of magnetic fields here is meant to emphasize the efficacy of our warped coordinates in maintaining stable circularly orbiting distributions of gas. 

\begin{figure}[htbp]
\centering
\includegraphics[width=0.6\textwidth]{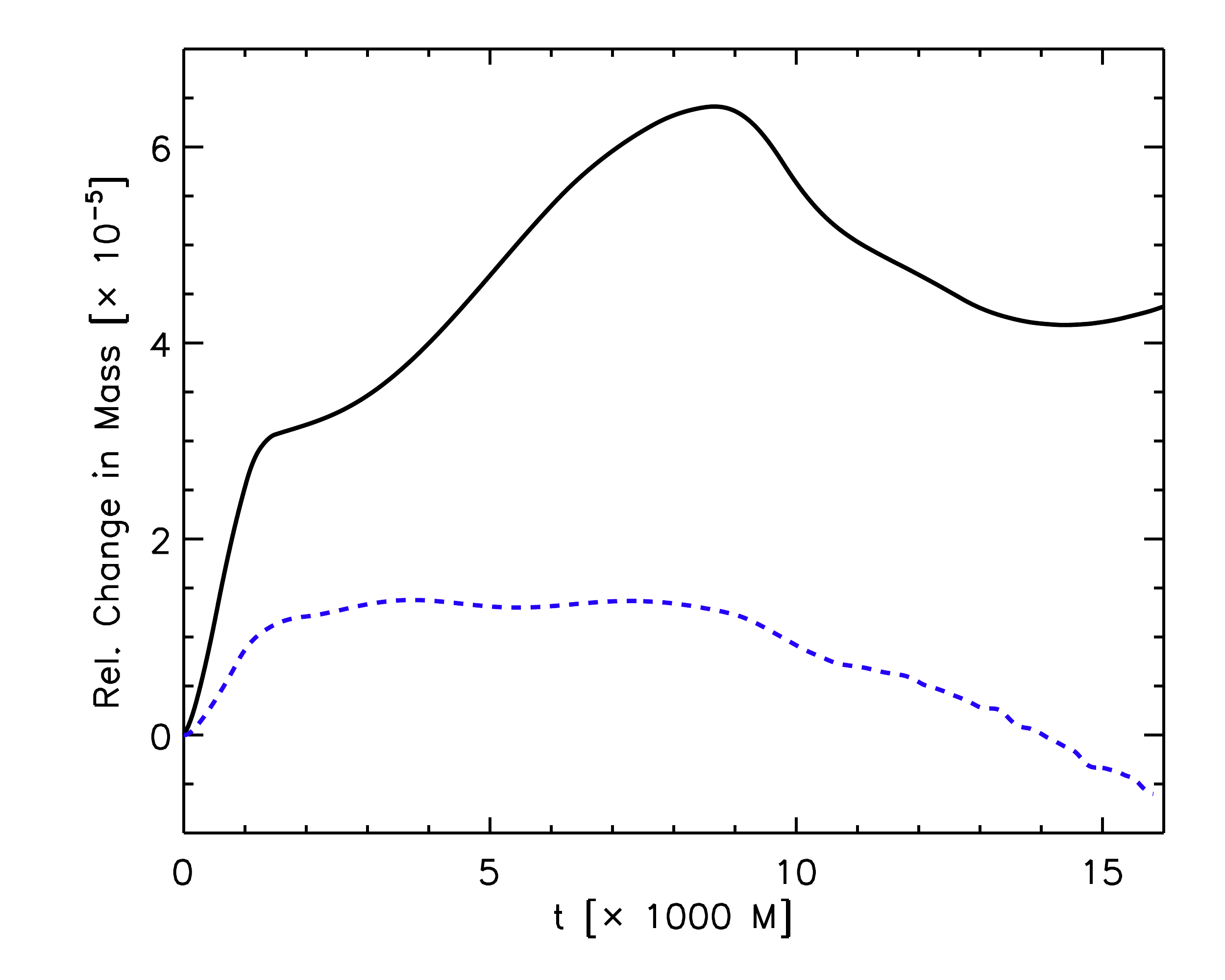}
\caption[]{Relative change in rest mass enclosed in the initial volume of the disk. We plot the result obtained with the warped grid configuration of Figure~\ref{fig:bbh-grid} (dashed blue line) and with the standard $\phi$-independent 
coordinate system of \harm (solid black line). \label{fig:diffusion} }
\end{figure}

We have further measured the diffusion of mass from the disk by computing the relative change in rest mass enclosed in the initial volume of the disk. The result obtained can be seen in Figure~\ref{fig:diffusion}. 
We compare the warped system's performance to that of an unwarped system,  which has been well tested and used
in many production runs in \harm.  The unwarped coordinates used in this test are defined in the 
following way:
$\xp{3} = \phi$, $\Delta \phi = 2 \pi / 400$, 
$\xp{1} = \ln(r) = \xp{1}_0 + i \Delta \xp{1}$, $\xp{1}_0 = \ln\left(\rin\right)$, 
$\Delta \xp{1} = \ln\left(\rout / \rin\right)/400$, $\rin = 1.845M$, and $\rout=300M$.    

After about $10^4 M$ or $6.3$ orbits at $r=\rin$, the 
dynamic coordinates have perturbed the inner edge enough so that a steady, though relatively weak, flow of material develops from the disk's edge.  This rate of inflow is smaller 
than the rate of mass increase
the two simulations experienced at the very beginning of their runs; the secular late time mass loss of the warped run is  $1/4$ times that 
of the warped run's initial mass increase and  $1/12$ times that of the unwarped run's.
The absorption of mass in the two runs is due to the initial data settling about a slightly different equilibrium state after evolution begins that is 
set by the truncation level of the EOM.  We also see larger swings in mass in the unwarped case than the rate of mass loss in the warped case.  The explanation for the poorer performance of the unwarped
run is that more of its resolution is focused near the horizon and not through the disk in comparison.  For instance, the unwarped radial grid spacing is about twice that of the warped case at the outer edge of the disk.


\subsection{Disk with black hole binary}
\label{sec:disk-BBH}

Finally, we present preliminary results for an evolution of an actual circumbinary accretion disk, which was the main motivation for this work.
For our analytic binary black hole spacetime model, we use a patchwork system of various approximate spacetime metrics (i.e., boosted black hole, post-Newtonian, and post-Minkowski metrics) 
to cover the entire physical domain,  and neglect the effect of the disk on the binary's orbit.  
The construction of the metric, initial data, and evolution procedure for this configuration is detailed in~\cite{Noble:2012xz,Mundimetal}, to where we refer the interested reader.  The only difference 
here is the disk is not magnetized and the disk lies only in the equator (i.e.\ the disk is assumed to be unstratified and is resolved by only one cell in the vertical dimension). 
Our test starts with two equal mass black holes at a fixed separation of $20~M$, $M \equiv M_{\rm BH1}+M_{\rm BH 2}$. The grid is one we have used before, namely that described in Figure~\ref{fig:bbh-grid}.  This grid is constructed so that there are 
approximately $32$ cells spanning each black hole horizon in each dimension. The distortions track the black holes on their orbits.  

Since the black holes now reside on the numerical domain, we ``excise'' a small volume of cells about each black hole's singularity from the update procedure to maintain a stable hydrodynamic evolution. 
The innermost metrics used to cover the spacetime about each black hole are written in horizon-penetrating coordinates, which allows us to excise cells well within the horizon.
Our excision method involves treating cells in three different ways depending on their proximity to the black holes' centers.  So-called ``excised'' cells are not evolved, and primitive variables and 
fluxes are not calculated at their faces.  There are also regular or ``evolved'' cells that are updated in the usual way and are shielded from the excised cells by so-called ``buffer'' cells.  The 
buffer cells ensure that evolved cells never use any data from excised cells, e.g., for the primitive variable reconstruction, the flux calculation, or the FluxCT procedure.  The buffer cells are not 
updated, but primitive variables are reconstructed and fluxes calculated at their faces.  In the test presented here, a cell is excised if any part of it lies within half a black hole's horizon radius from 
the black hole's center.  Since the stencil for our finite volume and FluxCT procedures span three cells on each side of a cell, buffer cells extend three cells in each dimension from the surface of the excised 
region.  The remaining cells are evolved cells.  

\begin{figure}[tbhp]
\centering
\includegraphics[width=0.7\textwidth]{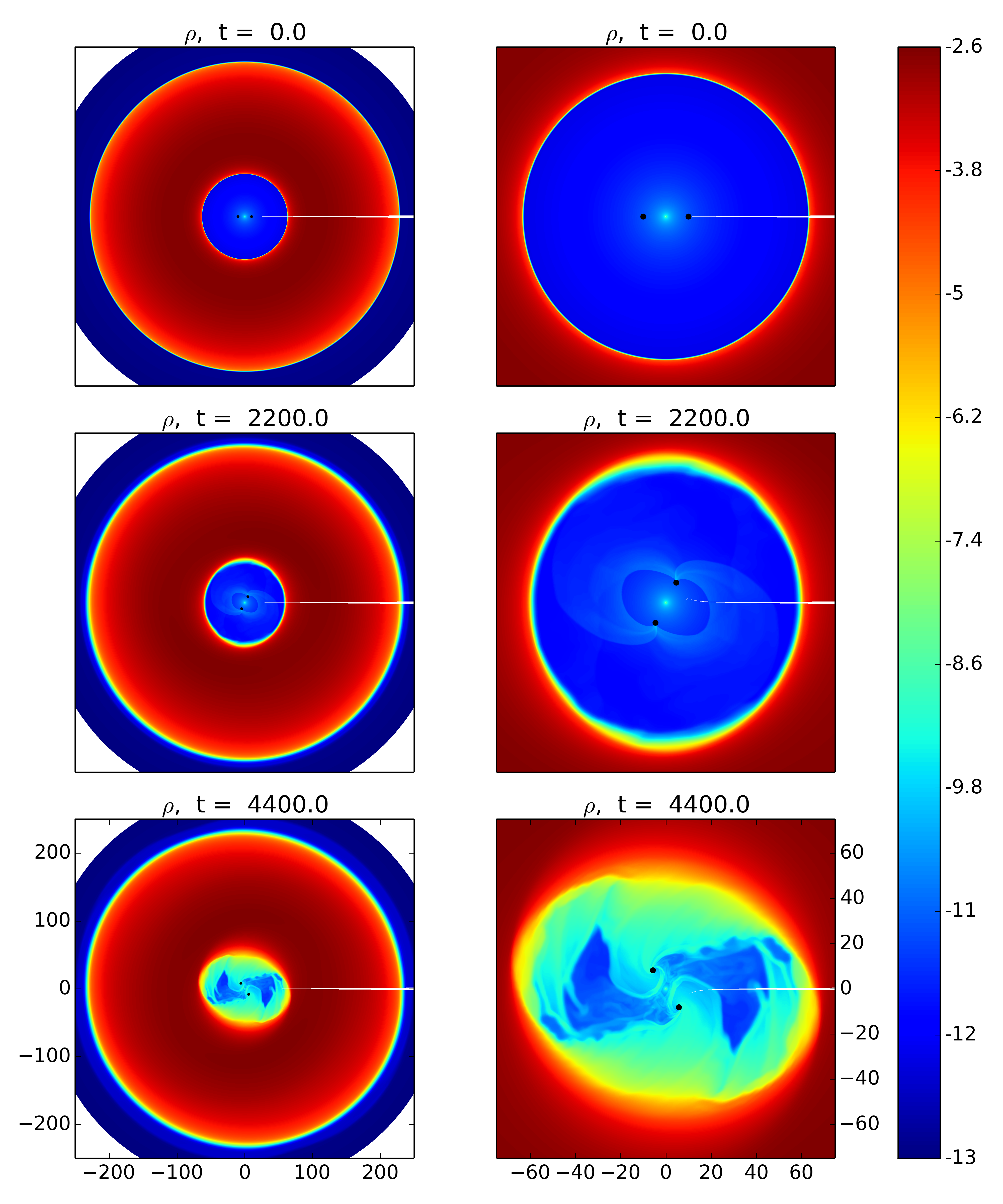}
\caption[]{Plotted is the density of the fluid, $\rho$, (in logarithmic scale) for a 2-d equatorial evolution of an accretion disk in the background of a black hole binary at $t=0$, $t=2200M$ and $t=4400M$. The black circles depict the black holes' horizons. We are here omitting the grid lines so as to not clutter the figure, but we note that the grid has the same configuration as the one presented in Figure~\ref{fig:bbh-grid}.  
The narrow white wedge appearing in all plots is a visualization artifact that marks the edge between the beginning and end of $\xp{3}$. 
\label{fig:BBH-disk-1} }
\end{figure}

In Figure~\ref{fig:BBH-disk-1} we can see the density of the fluid for three different time steps, $t=0,2200M,4400M$; these times occur after approximately $0, 4, 8$ orbits, respectively. Overdense regions of densities just above the atmosphere level quickly arise, once the 
simulation begins, trailing each black hole.  The overdensities are due to the fact that the numerically imposed floor or atmosphere is not in equilibrium with the binary's potential, allowing 
matter---which is artificially fed into the domain from the floor---to condense about each black hole.  This phenomenon is similar to what is seen with binaries in a similar environment---that of a nonrotating cloud
of gas~\cite{Bode:2009mt}.  
By $t=2200M$, the inner edge of the disk has begun to noticeably respond to the time-varying quadrupolar gravitational potential and distort from its original circular configuration.  

Very little accretion from the disk 
occurs until $t\simeq4000M$, after which gravitational torques begin to more efficiently draw material in toward the binary and dense accretion streams form from the disk to each black hole. 
The presence of these accretion streams is characteristic of circumbinary disk evolutions~\cite{Hayasaki07,Cuadra09,Farris11,Bode12,2012PhRvL.109v1102F}.  One key difference we find is
that the accretion streams seem to exhibit the Kelvin-Helmholtz instability, recognizable by the turbulent eddies and waves that form along the shear layer at the edge of the streams. 
As far as the authors know, the Kelvin-Helmholtz instability has not been observed before in the context of circumbinary accretion onto black holes. 
We suppose that we may be seeing it because of the higher effective accuracy of our hydrodynamics methods.  We intend to further investigate the accretion streams and
verify that this is in fact the Kelvin-Helmholtz instability in future work. 

We find that the warped coordinates sufficiently resolve material flowing near each black hole and that our excision procedure works as desired.  Material flows into each black hole without 
reflection, and no hydrodynamic waves appear to emanate from them.   The warped grid also conforms to the symmetry of the inner edge of the disk well enough so that accretion occurs later in time, on the 
timescale set by the binary's gravitational torques, and not by the effective grid scale viscosity present in---for instance---Cartesian coordinates.
Future efforts include exploring the role the Kelvin-Helmholtz instability plays in the evolution of the streams, and incorporating  magnetic fields in 3-d simulations using these coordinates. 

\section{Final remarks}
\label{sec:final}

In this paper, we have constructed a warped gridding scheme adapted to binary black hole simulations
and discussed its implementation in the GRMHD \harm code.   

In order to demonstrate the coordinate system's efficacy and to validate its implementation a number of tests were performed.  We have shown results obtained for the advection of a magnetic field loop, 
the evolution of the Bondi solution, and the evolution of a stationary gaseous disk in the background of a single black hole.  We lastly presented preliminary 
results for the evolution of a circumbinary disk, the main motivation for this work.
The warped coordinate evolution in the advected field loop test performed even better than the uniform Cartesian system, demonstrating 
that the distortions do not lead to extra diffusion in the magnetic field evolution.   The dynamic and asymmetric aspect of the coordinates does introduce additional 
errors at the truncation level when compared to symmetric grids and solutions (e.g., the Bondi solution, stationary disk), but these errors do diminish with increasing 
cell count and are much smaller than what we would find if we were to use Cartesian grids with the same cell count.  For instance, the stationary disk solution
accreted no mass until after tens of binary orbits and several orbits at the disk's inner edge---whereas disk evolutions on Cartesian grids immediately start accreting gas. 
Also, our results demonstrated that the distortions in the grid were able to resolve small scale features as expected 
and produced little---if any---anomalous features in the solutions.  

We note that, while our results were very good for the warped grid configurations we tested, they do not necessarily extrapolate to different grid configurations. Indeed, we observed poorer results when the grid chosen was significantly distorted and not at all adapted to the symmetries of the problem.  Care must therefore be taken when choosing the grid parameters, and some degree of testing and experimentation should be expected when using this construction.

Another consideration to make is one of computational cost.   
While the evaluation of all the functions related to the coordinate transformation carries some overhead in the code, 
we mitigated this impact by eliminating redundancies by storing auxiliary terms in memory rather than recalculating them.  
In our case, our spacetime metric is specified in Cartesian coordinates, so a coordinate transformation to spherical coordinates
is necessary regardless of whether we use the coordinates described here.  

We stress that our construction and resulting grid are sufficiently general that they can be applied to very different problems and codes.
As are many other GRMHD codes (e.g.,~\cite{Moesta:2013dna}), \harm is written in a covariant way making the adoption of this coordinate system straightforward.  
For instance, it would be interesting to see if dynamic gauge conditions used to evolve punctures, would be stable under these dynamic coordinate changes.   
Our approach could be especially interesting for modular unigrid codes as a simpler alternative to AMR, particular for problems tracking compact features.   
The satisfactory results presented here bring us high hopes for ongoing efforts towards modeling circumbinary accretion flows.


\ack
We appreciate the helpful input from our collaborators M. Campanelli, Y. Zlochower, J. Krolik, C. Lousto, B. Mundim, H. Nakano, and K. Sorathia. 
M.Z. is supported by NSF grants OCI-0832606, PHY-0969855, AST-1028087, and PHY-1229173.
S.C.N. is supported by NSF grant No. OCI-0725070, OCI-0832606, AST-1028087, PHY-1125915. 
Computational resources were provided by XSEDE allocation TG-PHY060027N, and by NewHorizons and BlueSky Clusters at Rochester Institute of Technology, which were supported by NSF grant No. PHY-0722703, DMS-0820923, AST-1028087, and PHY-1229173. 

\appendix

\section{Magnetic Fields on Generally Warped Coordinates}
\label{app:magn-fields-gener}

It is often the case that initial data routines assume properties of the numerical 
coordinates on which they reside.  For instance, many initial data distributions are 
constant with respect to the azimuthal coordinate---which is often assumed to be the 
same as one of the numerical coordinates---so they tend to copy the data along this 
azimuthal array dimension, or---for instance---use this independence to derive the 
magnetic field in a simple way.   When using warped coordinates, however, we can no 
longer make such simplifying assumptions and must calculate the initial data in the most general
way.  We have made these changes to the relevant initial data routines used by \harm so 
that assumptions on the coordinate system no longer remain.  In the following, 
we describe the new method for initializing the magnetic field.  

First, let us present a short review of classical electrodynamics (please see 
\cite{1962clel.book.....J,2003ApJ...585..921B} for more details).  Using the Faraday 
tensor, $F^{\mu \nu}$, and the Maxwell tensor, $\dF^{\mu \nu}$, we can succinctly express the 
Maxwell equations as 
\beq{
\nabla_\mu F^{\nu \mu} = J^\nu \quad , \label{maxwell-eqs1}
}
\beq{
\nabla_\mu \dF^{\mu \nu} = 0 \quad . \label{maxwell-eqs2}
}
We have absorbed factors of $\left(1/\sqrt{4 \pi}\right)$ into the tensors for the sake of simplicity.   
Note that $F^{\mu \nu}$ and $\dF^{\mu \nu}$ 
are both antisymmetric, e.g. $F^{\mu \nu}=-F^{\nu \mu}$ and $\dF^{\mu \nu}=-\dF^{\nu \mu}$. 
The first set yields the 
evolution equations for the electric field, while the second set gives rise to the induction equations 
and the solenoidal constraint.  The electromagnetic part of the stress-energy tensor is
\beq{
T^{\mu\nu}_\mathrm{EM} = F^{\mu \lambda} {F^\nu}_\lambda 
- \frac{1}{4} g^{\mu \nu} F^{\lambda \kappa}  F_{\lambda \kappa} 
= \bsq u^\mu u^\nu + \frac{1}{2} \bsq g^{\mu\nu} - b^\mu b^\nu \quad 
 , \label{em-stress}
}
where the magnetic 4-vector is $b^\mu = \dF^{\nu \mu} u_\nu$, 
and $\dF^{\mu \nu} = b^\mu u^\nu - b^\nu u^\mu$.  We typically use the spatial magnetic field vector, 
\beq{
B^i = \frac{1}{\alpha} \sB^i  = \dF^{it} \label{spatial-mag-vector}
}
where 
\beq{
\sB^\mu = n_\nu \dF^{\nu \mu} \label{mag-vector}
}
and $n_\mu = [-\alpha,0,0,0]$, $n^\mu = \frac{1}{\alpha}[1,-\beta^i]$. 
One can easily show that the $i$-component of Eq.~(\ref{maxwell-eqs2}) gives rise to the 
induction equation:
\beq{
\partial_t \sqrt{-g} B^i + \partial_j \sqrt{-g} \left(b^i u^j - b^j u^i\right)  = 0 
\label{induction-eq}
}
and the solenoidal constraint (aka ``divergence-less constraint'', aka ``div.b=0'' condition, aka ``no magnetic monopoles'' condition, etc.):
\beq{
\partial_i \sqrt{-g} B^i = 0  \quad . \label{solenoidal-constraint}
}
Note that this comes from the time component of Eq.~(\ref{maxwell-eqs2}):
\beqa{
0 & = & \nabla_\mu \dF^{\mu t}  =  \frac{1}{\sqrt{-g}} \left[ \partial_\mu \left( \sqrt{-g} \ \dF^{\mu t} \right)  + \sqrt{-g} {\Gamma^{t}}_{\mu \kappa} \dF^{\mu \kappa} \right] \label{divb-deriv1}\\
  & = & \frac{1}{\sqrt{-g}} \left[ \partial_i \sqrt{-g} B^i \right] \label{divb-deriv5}
}
where Eq.~(\ref{divb-deriv5}) follows Eq.~(\ref{divb-deriv1}) because $\dF^{\mu \nu}$ is antisymmetric 
and the Christoffel symbol is symmetric in the lower indices.   
Note that this is a gauge-independent constraint, i.e.\ one can perform an arbitrary coordinate 
transformation at the beginning and end up with a divergence-less quantity in that new coordinate system.   All that is required is that both 
the derivative operator and the magnetic field vector be in the \emph{same} coordinate system.   Since our finite difference 
operators are w.r.t.\ the uniform numerical coordinates, then $B^i$ must be in numerical coordinates as well. 

Finite difference solutions are only accurate to truncation error.  Different stencils---or finite difference approximations---for the divergence operator of the 
solenoidal constraint give rise to different truncation errors.  The idea behind the CT method is that one 
special stencil for the difference operation is chosen so that the solenoidal constraint is satisfied to machine precision. 
This special stencil or finite difference operator is called ``compatible'' with the CT scheme. The finite difference operator for first-order spatial derivatives
compatible with our FluxCT scheme is:
\beq{
\fl \hat{\Delta}_{\xp{1}} B_{[l_1-1/2,l_2-1/2,l_3-1/2]} = \frac{1}{4\dxp{1}} \sum_{p = -1}^{0} \sum_{q = -1}^{0} 
\left( B_{[l_1,l_2+p,l_3+q]} - B_{[l_1-1,l_2+p,l_3+q]} \right)
\label{flux-ct-stencil-x1}
}
with similar equations for the $\hat{\Delta}_{\xp{2}}$ and $\hat{\Delta}_{\xp{3}}$ operators, and 
where $B_{[i,j,k]}$ is the value of a magnetic field component at the numerical coordinates of the center of the cell with spatial indices $\left\{i,j,k\right\}$. 
Our compatible stencil is centered at the cell corners while our magnetic 
field vectors are cell-centered quantities.  One can describe the compatible stencil as a corner-centered difference of edge-centered quantities, which are 
found by averaging the four nearest cell-centered values to the edge centers.  To be explicit, the following divergence operation is preserved to round-off errors in our FluxCT scheme:
\beq{
\mathbf{\nabla}_{\mathrm{CT}} \cdot \mathbf{B} \ = \  \left[ \hat{\Delta}_{\xp{1}}  B^1 \ + \ \hat{\Delta}_{\xp{2}} B^2 \ + \ \hat{\Delta}_{\xp{3}} B^3 \right]_{[l_1-1/2,l_2-1/2,l_3-1/2]} \label{fluxct-divb}
}

In order to initialize the magnetic field such that it is divergence-less w.r.t.\ this stencil, 
we set the vector potential at the cell corners and take the curl of it using the same derivative operators 
as were used for $\mathbf{\nabla}_{\mathrm{CT}}$.  We remind the reader that the covariant $4$-vector potential
is $\mathcal{A}_\mu =  \Phi n_\mu + A_\mu$, where $\Phi$ is the electrostatic scalar potential, and $A_\mu$ is the 
spatial vector potential, where $A_\mu n^\mu = 0$, $A_t = \beta^i A_i$, $A^t = 0$, and $\mathcal{A}_i = A_i$~\cite{1962clel.book.....J,2003ApJ...585..921B}.  
The magnetic field is calculated from the curl of the vector potential:
\beq{
B^i = \frac{1}{\sqrt{-g}} \, \tilde{\epsilon}^{ijk} \, \partial_j A_k    \quad , \label{curl-of-A}
}
where 
\beq{
\tilde{\epsilon}^{ijk} = \sqrt{\gamma} \, \epsilon^{ijk} = \sqrt{\gamma} \, n_\mu \epsilon^{\mu ijk}  = - \sqrt{-g} \, \epsilon^{tijk}  
\quad . \label{epsilon}
}
and $\tilde{\epsilon}^{ijk}$ is the anti-symmetric permutation tensor that equals $1$ for even permutations of $(1,2,3)$, 
$-1$ for odd permutations, and zero for repeating indices.  Also, here, $\epsilon^{\mu \nu \kappa \lambda}$ and  $\epsilon^{i j k}$ are, respectively, the standard 
$4$-dimensional and $3$-dimensional Levi-Civita symbols.  It is tedious, though straightforward, to 
show that 
\beq{ 
B^i_{[l_1,l_2,l_3]} = \left[ \frac{1}{\sqrt{-g}} \, \tilde{\epsilon}^{ijk} \, \hat{\Delta}_{\xp{j}} A_k \right]_{[l_1,l_2,l_3]}   
\quad \Rightarrow \quad \left[ \mathbf{\nabla}_{\mathrm{CT}} \cdot \mathbf{B} \right]_{[l_1,l_2,l_3]} = 0  \quad .  \label{divergence-less-stencil-condition}
}
We typically set $A_\phi$, transform $A_i$ to numerical coordinates, and evaluate the curl in numerical coordinates using $\hat{\Delta}_{\xp{j}}$ to approximate the 
derivatives in the curl expression.  We need not calculate the $\mathcal{A}_t$ component before transforming since $dt/dx^{\prime i} = 0$ for the coordinates of 
interest here. 

In order to enforce the divergenceless constraint, we use a 3-d version of the FluxCT algorithm described in~\cite{GMT03,Toth00}.  The principal idea behind most 
CT schemes is that a consistent set of EMFs are used to reconstruct the fluxes in the induction equation so that, when the 
discretized time derivative and the CT-consistent divergence operator both act on the magnetic field, the result is zero algebraically.  Please 
see~\cite{1988ApJ...332..659E} for a thorough description of how CT schemes are constructed, as we will only describe here what is necessary for reproducing our scheme. 
The first step in our procedure is to reconstruct the primitive variables and calculate the numerical fluxes at each cell face.  It is easy to see that the 
spatial fluxes of the induction equation are components of the Maxwell tensor: 
\beq{
\partial_t \sqrt{-g} B^i + \partial_j \left( \sqrt{-g} \ \dF^{ij} \right)  = 0  \quad . 
\label{induction-eq-2}
}
If a cell's center is located at the point with indices $[l_1, l_2, l_3]$, let that cell's faces in the $\xp{i}$ direction be offset by $\pm 1/2$ in that dimension, e.g.,
the faces orthogonal to $\xp{1}$ have indices $[l_1 \pm 1/2, l_2, l_3]$.   The cell's edges along that face are then half intervals in the other dimensions, e.g., for 
the top $\xp{1}$ face the indices are $\left\{ [l_1+1/2, l_2 \pm 1/2 , l_3 ], [l_1+1/2, l_2  , l_3 \pm 1/2] \right\}$.  We next use the numerical fluxes from the first step to 
interpolate for a common set of edge-centered EMFs shared by all cells.   This procedure exploits the analytic identity that $\emf{k} = - \tilde{\epsilon}_{i j k} b^i u^j$ and 
$\dF^{i j} = - \tilde{\epsilon}^{i j k} \emf{k}$.  For each EMF component, an average is taken between two interpolated quantities, with each interpolation performed along a 
different dimension.  Specifically, the EMFs are: 
\beqa{
\emf{1}_{[l_1       , l_2 - 1/2 , l_3 - 1/2]}  & = & \frac{1}{2} \left[ \left< \dFn^{3 2} \right>_{[l_1       , l_2 - 1/2 , l_3 - 1/2]}  -  \left< \dFn^{2 3} \right>_{[l_1       , l_2 - 1/2 , l_3 - 1/2]}  \right], \label{emf-1} \\[0.25cm]
\emf{2}_{[l_1 - 1/2 , l_2       , l_3 - 1/2]}  & = & \frac{1}{2} \left[ \left< \dFn^{1 3} \right>_{[l_1 - 1/2 , l_2       , l_3 - 1/2]}  -  \left< \dFn^{3 1} \right>_{[l_1 - 1/2 , l_2       , l_3 - 1/2]}  \right], \label{emf-2} \\[0.25cm]
\emf{3}_{[l_1 - 1/2 , l_2 - 1/2 , l_3      ]}  & = & \frac{1}{2} \left[ \left< \dFn^{2 1} \right>_{[l_1 - 1/2 , l_2 - 1/2 , l_3      ]}  -  \left< \dFn^{1 2} \right>_{[l_1 - 1/2 , l_2 - 1/2 , l_3      ]}  \right], \label{emf-3} 
}
where $\dFn^{i j}$ are the numerical fluxes calculated in the first step, and 
the angle brackets denote that the interior quantity has been interpolate to the indicated position.  For the ``linear'' FluxCT procedure, we use centered first-order interpolation, equivalent 
to nearest neighbor averaging: 
\beqa{
\left< \dFn^{1 2} \right>_{[l_1 - 1/2 , l_2 - 1/2 , l_3      ]} & = & \frac{1}{2} \left[ \dFn^{1 2}_{[l_1 , l_2 - 1/2 , l_3 ]} + \dFn^{1 2}_{[l_1 - 1 , l_2 - 1/2 , l_3 ]} \right], \label{dfn-12} \\
\left< \dFn^{2 1} \right>_{[l_1 - 1/2 , l_2 - 1/2 , l_3      ]} & = & \frac{1}{2} \left[ \dFn^{2 1}_{[l_1 - 1/2 , l_2 , l_3 ]} + \dFn^{2 1}_{[l_1 - 1/2 , l_2 - 1 , l_3 ]} \right], \label{dfn-21} \\
\left< \dFn^{1 3} \right>_{[l_1 - 1/2 , l_2       , l_3 - 1/2]} & = & \frac{1}{2} \left[ \dFn^{1 3}_{[l_1 , l_2 , l_3 - 1/2 ]} + \dFn^{1 3}_{[l_1 - 1 , l_2 , l_3 - 1/2 ]} \right], \label{dfn-13} \\
\left< \dFn^{3 1} \right>_{[l_1 - 1/2 , l_2       , l_3 - 1/2]} & = & \frac{1}{2} \left[ \dFn^{3 1}_{[l_1 - 1/2 , l_2 , l_3 ]} + \dFn^{3 1}_{[l_1 - 1/2 , l_2 , l_3 - 1 ]} \right], \label{dfn-31} \\
\left< \dFn^{2 3} \right>_{[l_1       , l_2 - 1/2 , l_3 - 1/2]} & = & \frac{1}{2} \left[ \dFn^{2 3}_{[l_1 , l_2 , l_3 - 1/2 ]} + \dFn^{2 3}_{[l_1 , l_2 - 1 , l_3 - 1/2 ]} \right], \label{dfn-23} \\
\left< \dFn^{3 2} \right>_{[l_1       , l_2 - 1/2 , l_3 - 1/2]} & = & \frac{1}{2} \left[ \dFn^{3 2}_{[l_1 , l_2 - 1/2 , l_3 ]} + \dFn^{3 2}_{[l_1 , l_2 - 1/2 , l_3 - 1 ]} \right]. \label{dfn-32} 
}
For the ``parabolic'' FluxCT scheme, the interpolation is performed using the same piecewise parabolic interpolation method we use for reconstructing face-centered quantities when computing the numerical fluxes. 
Lastly, the ultimate divergenceless-preserving induction equation fluxes ($\dFs^{ij}$) are calculated by linearly interpolating a pair of edge EMFs to the face's center:
\beqa{
\dFs^{1 2}_{[l_1 , l_2 - 1/2 , l_3 ]}  & = -&\frac{1}{2} \left[ \emf{3}_{[l_1 - 1/2 , l_2 - 1/2 , l_3 ]}  + \emf{3}_{[l_1 + 1/2 , l_2 - 1/2 , l_3 ]}  \right], \label{dfs-12} \\
\dFs^{2 1}_{[l_1 - 1/2 , l_2 , l_3 ]}  & =  &\frac{1}{2} \left[ \emf{3}_{[l_1 - 1/2 , l_2 - 1/2 , l_3 ]}  + \emf{3}_{[l_1 - 1/2 , l_2 + 1/2 , l_3 ]}  \right], \label{dfs-21} \\
\dFs^{1 3}_{[l_1 , l_2 , l_3 - 1/2 ]}  & =  &\frac{1}{2} \left[ \emf{2}_{[l_1 - 1/2 , l_2 , l_3 - 1/2 ]}  + \emf{2}_{[l_1 + 1/2 , l_2 , l_3 - 1/2 ]}  \right], \label{dfs-13} \\
\dFs^{3 1}_{[l_1 - 1/2 , l_2 , l_3 ]}  & = -&\frac{1}{2} \left[ \emf{2}_{[l_1 - 1/2 , l_2 , l_3 - 1/2 ]}  + \emf{2}_{[l_1 - 1/2 , l_2 , l_3 + 1/2 ]}  \right], \label{dfs-31} \\
\dFs^{2 3}_{[l_1 , l_2 , l_3 - 1/2 ]}  & = -&\frac{1}{2} \left[ \emf{1}_{[l_1 , l_2 - 1/2 , l_3 - 1/2 ]}  + \emf{1}_{[l_1 , l_2 + 1/2 , l_3 - 1/2 ]}  \right], \label{dfs-23} \\
\dFs^{3 2}_{[l_1 , l_2 - 1/2 , l_3 ]}  & =  &\frac{1}{2} \left[ \emf{1}_{[l_1 , l_2 - 1/2 , l_3 - 1/2 ]}  + \emf{1}_{[l_1 , l_2 - 1/2 , l_3 + 1/2 ]}  \right]. \label{dfs-32} 
}
These fluxes are those used to update the magnetic field components.  


\bibliography{bhm_references,references}

\end{document}